\documentclass[nojss]{jss}
\usepackage{amsmath}
\usepackage{graphicx}%
\usepackage{amsfonts}%
\usepackage{amssymb}
\usepackage{overpic}
\usepackage{subfigure}
\usepackage{rotating}
\usepackage{url}
\usepackage{textcomp}
\usepackage{lineno}
\usepackage{paralist}
\usepackage{setspace}
\usepackage{tabularx}
\usepackage{natbib}

\newcommand{\bA}{ {\boldsymbol A} }
\newcommand{\bb}{ {\boldsymbol b} }
\newcommand{\bB}{ {\boldsymbol B} }
\newcommand{\bc}{ {\boldsymbol c} }
\newcommand{\bC}{ {\boldsymbol C} }

\newcommand{\bD}{ {\boldsymbol D} }

\newcommand{\bG}{ {\boldsymbol G} }

\newcommand{\bH}{ {\boldsymbol H} }

\newcommand{\bI}{ {\boldsymbol I} }

\newcommand{\bK}{ {\boldsymbol K} }

\newcommand{\bL}{ {\boldsymbol L} }
\newcommand{\bm}{ {\boldsymbol m} }
\newcommand{\bM}{ {\boldsymbol M} }

\newcommand{\bO}{ {\boldsymbol O} }

\newcommand{\bs}{ {\boldsymbol s} }

\newcommand{\bt}{ {\boldsymbol t} }
\newcommand{\bT}{ {\boldsymbol T} }
\newcommand{\bu}{ {\boldsymbol u} }
\newcommand{\bU}{ {\boldsymbol U} }
\newcommand{\bv}{ {\boldsymbol v} }
\newcommand{\bV}{ {\boldsymbol V} }
\newcommand{\bw}{ {\boldsymbol w} }
\newcommand{\bW}{ {\boldsymbol W} }
\newcommand{\bx}{ {\boldsymbol x} }
\newcommand{\bX}{ {\boldsymbol X} }
\newcommand{\by}{ {\boldsymbol y} }

\newcommand{\bz}{ {\boldsymbol z} }
\newcommand{\bZ}{ {\boldsymbol Z} }

\newcommand{\balpha}{ {\boldsymbol \alpha} }

\newcommand{\bbeta}{ {\boldsymbol \beta} }

\newcommand{\bepsilon}{ {\boldsymbol \varepsilon} }
\newcommand{\bphi}{ {\boldsymbol \phi} }

\newcommand{\bmu}{ {\boldsymbol \mu} }

\newcommand{\bet}{ {\boldsymbol \eta} }

\newcommand{\bSigma}{ {\boldsymbol \Sigma} }

\newcommand{\btheta}{ {\boldsymbol \theta} }

\newcommand{\bPsi}{ {\boldsymbol \Psi} }

\newcommand{\bzero}{ {\boldsymbol 0} }

\newcommand{\given}{\,|\,}

\newcommand{\tildew}{\tilde{w}}
\newcommand{\tildebw}{\tilde{\bw}}

\newcommand{\R}{ \proglang{R}}

\newcommand{\chol}{\mbox{\texttt{chol}}}
\newcommand{\trsolve}{\mbox{\texttt{trsolve}}}

\author{Andrew O. Finley\\Michigan State University \And 
        Sudipto Banerjee\\University of Minnesota \And 
        Alan E. Gelfand\\Duke University}
\title{\pkg{spBayes} for large univariate and multivariate point-referenced spatio-temporal data models}

\Plainauthor{Andrew O. Finley, Sudipto Banerjee, Alan E.Gelfand} 
\Plaintitle{\pkg{spBayes} for large univariate and multivariate point-referenced spatio-temporal data models} 
\Shorttitle{\pkg{spBayes} point-referenced spatio-temporal data models} 

\Abstract{
  In this paper we detail the reformulation and rewrite of core functions in the \pkg{spBayes} {\R} package. These efforts have focused on improving computational efficiency, flexibility, and usability for point-referenced data models. Attention is given to algorithm and computing developments that result in improved sampler convergence rate and efficiency by reducing parameter space; decreased sampler run-time by avoiding expensive matrix computations, and; increased scalability to large datasets by implementing a class of \emph{predictive process} models that attempt to overcome computational hurdles by representing spatial processes in terms of lower-dimensional realizations. Beyond these general computational improvements for existing model functions, we detail new functions for modeling data indexed in both space and time. These new functions implement a class of dynamic spatio-temporal models for settings where space is viewed as continuous and time is taken as discrete.}

\Keywords{spatial, temporal, multivariate, Gaussian predictive process, Markov chain Monte Carlo}
\Plainkeywords{spatial, temporal, multivariate, Gaussian predictive process, Markov chain Monte Carlo} 

\Address{
  Andrew O. Finley\\
  Departments of Forestry and Geography\\
  Michigan State University\\
  Natural Resources Building\\
  480 Wilson Road, Room 126\\
  East Lansing, MI 48824-1222\\
  E-mail: \email{finleya@msu.edu}\\
  URL: \url{http://blue.for.msu.edu}
}

\begin{document}

\section{Introduction}\label{intro}
The scientific community is moving into an era where open-access data-rich environments provide extraordinary opportunities to understand the spatial and temporal complexity of processes at broad scales. Unprecedented access to spatial data is a result of investments to collect data for regulatory, monitoring, and resource management objectives, and technological advances in spatially-enabled sensor networks along with geospatial information storage, analysis, and distribution systems. These data sources are increasingly diverse and specialized, e.g., computer model outputs, monitoring station instruments, remotely located sensors, and georeferenced field measurements. Across scientific fields, researchers face the challenge of coupling these data with imperfect models to better understand variability in their system of interest. The inference garnered through these analyses often supports decisions with important economic, environmental, and public health implications; therefore, it is critical to correctly estimate inferential uncertainty. However, developing modeling frameworks capable of accounting for various sources of uncertainty is not a trivial task---massive datasets from multiple sources with complex spatial dependence structures only serve to aggravate the challenges. 

Proliferation of spatial data has spurred considerable development in statistical modeling; see, for example, the books by \cite{cressie1993}, \cite{chil2012}, \cite{moller2003}, \cite{schabenberger2004}, \cite{wackernagel2003}, \cite{diggle2007} and \cite{cressie2011} for a variety of methods and applications. The statistical literature acknowledges that spatial and temporal associations are captured most effectively using models that build dependencies in different stages or hierarchies. Hierarchical models are especially advantageous with datasets having several lurking sources of uncertainty and dependence, where they can estimate much richer models with less stringent assumptions than traditional modeling paradigms. These models follow the Bayesian framework of statistical inference (see, e.g., \citealt{carlin2011}; \citealt{gelman2004}), where analysis uses sampling from the posterior distributions of model parameters. 

Computational advances with regard to Markov chain Monte Carlo (MCMC) methods have contributed enormously to the popularity of hierarchical  models in a wide array of disciplines (e.g., \citealt{gilks1996}; \citealt{robert2010}), and spatial modeling is no exception (see, e.g., \citealt{banerjee2004}). In the realm of spatial statistics, hierarchical models have been widely applied to analyze both areally referenced as well as point-referenced or geostatistical data. For the former, a class of models known as Conditionally Autoregressive (CAR) models have become very popular as they are easily implemented using MCMC methods such as the Gibbs sampler. In fact, these models are somewhat naturally suited for the Gibbs sampler which draws samples from conditional distributions that are fully specified by the CAR models. Their popularity has increased in no small measure due to their automated implementation in the OpenBUGS software package which offers a flexible and user-friendly interface to construct multilevel models that are implemented using a Gibbs sampler. This is performed by identifying a multilevel model with a directed acyclic graph (DAG) whose nodes form the different components of the model and allow the language to identify the full conditional distributions that need to be updated. OpenBUGS is an offshoot of the BUGS (Bayesian inference Using Gibbs Sampling) project and the successor of the WinBUGS software. 

From an automated implementation perspective, the challenges are somewhat greater for point-referenced models. First, expensive matrix computations are required that can become prohibitive with large datasets. Second, routines to fit unmarginalized models are less suited for direct updating using a Gibbs sampler in the BUGS paradigm and results in slower convergence of the chains. Third, investigators often encounter multivariate spatial datasets with several spatially dependent outcomes, whose analysis requires multivariate spatial models that involve matrix computations that are poorly implemented in BUGS. These issues have, however, started to wane with the delivery of relatively simpler {\R} \citep{R} packages via the Comprehensive R Archive Network (CRAN) (\url{http://cran.r-project.org}) that help automate Bayesian methods for point-referenced data and diagnose convergence. The Analysis of Spatial Data \citep{spTV} and Handling and Analyzing Spatio-Temporal Data \citep{sptimeTV} CRAN Task Views provide a convenient way to identify packages that offer functions for modeling such data. These packages are generally listed under the \emph{Geostatisics} section in the Task View. Here, those packages that fit Bayesian model include \pkg{geoR} \citep{geoR}, \pkg{geoRglm} \citep{geoRglm}, \pkg{spTimer} \citep{geoRglm}, \pkg{spBayes} \citep{spBayes}, \pkg{spate} \citep{spate}, and \pkg{ramps} \citep{ramps}. In terms of functionality, \pkg{spBayes} offers users a suite of Bayesian hierarchical models for Gaussian and non-Gaussian univariate and multivariate spatial data as well as dynamic Bayesian spatial-temporal models.

Our initial development of \pkg{spBayes} \citep{finley2007} provided functions for modeling Gaussian and non-Gaussian univariate and multivariate point-referenced data. These hierarchical Bayesian spatial process models, implemented through MCMC methods, offered increased flexibility to fit models that would be infeasible with classical methods within inappropriate asymptotic paradigms. However, with this increased flexibility comes substantial computational demands. Estimating these models involves expensive matrix decompositions whose computational complexity increases in cubic order with the number of spatial locations, rendering such models infeasible for large spatial datasets. Through \pkg{spBayes} version 0.2-4, released on CRAN on 4/24/12, very little attention was given to addressing these computational challenges. As a result, fitting models with more than a few hundred observations was very time consuming---on the order of hours to fit models with $\sim$1,000 locations.

\pkg{spBayes} version 0.3-7 (CRAN 6/1/13) comprises a substantial reformulation and rewrite of core functions for model fitting, with a focus on improving computational efficiency, flexibility, and usability. Among other improvements, this and subsequent versions offer: $i$) improved sampler convergence rate and efficiency by reducing parameter space; $ii$) decreased sampler run-time by avoiding expensive matrix computations, and; $iii$) increased scalability to large datasets by implementing a class of \emph{predictive process} models that attempt to overcome computational hurdles by representing spatial processes in terms of lower-dimensional realizations. Beyond these general computational improvements for existing models, new functions were added to model data indexed in both space and time. These functions implement a class of dynamic spatio-temporal models for settings where space is viewed as continuous and time is taken as discrete. The subsequent sections highlight the fundamentals of models now implemented in \pkg{spBayes}. 

\section{Bayesian Gaussian spatial regression models}\label{Sec: Bayesian_Gaussian_Generic}

\cite{finley2007} outline the first version of \pkg{spBayes} as an {\R} package for estimating Bayesian spatial regression models for point-referenced outcomes arising from Gaussian, binomial or Poisson distributions. For the Gaussian case, the recent version of \pkg{spBayes} offers several Bayesian spatial models emerging from the hierarchical linear mixed model framework
\begin{align}\label{Eq: Bayesian_Spatial_Gaussian_Generic}
p(\btheta) \times N(\bbeta\given \bmu_{\beta}, \bSigma_{\beta})\times N(\balpha\given \bzero, \bK(\btheta)) \times N(\by \given \bX\bbeta + \bZ(\btheta)\balpha, \bD(\btheta))\; , 
\end{align}
where $\by$ is an $n\times 1$ vector of possibly irregularly located observations, $\bX$ is a known $n\times p$ matrix of regressors ($p < n$), $\bK(\btheta)$ and $\bD(\btheta)$ are families of $r\times r$ and $n\times n$ covariance matrices, respectively, and $\bZ(\btheta)$ is $n\times r$ with $r\leq n$, all indexed by a set of unknown process parameters $\btheta$. The $r\times 1$ random vector $\balpha\sim N(\bzero, \bK(\btheta))$ and the $p\times 1$ slope vector $\bbeta\sim N(\bmu_{\beta},\bSigma_{\beta})$, where $\bmu_{\beta}$ and $\bSigma_{\beta}$ are known. The hierarchy is completed by assuming $ \btheta \sim p(\btheta)$, a proper prior distribution. The Gaussian spatial models in \pkg{spBayes} emerge as special cases of (\ref{Eq: Bayesian_Spatial_Gaussian_Generic}), which we will see later. Bayesian inference is carried out by sampling from the posterior distribution of $\{\bbeta,\balpha,\btheta\}$, which is proportional to (\ref{Eq: Bayesian_Spatial_Gaussian_Generic}).

Below, we provide some details behind Bayesian inference for (\ref{Eq: Bayesian_Spatial_Gaussian_Generic}). This involves sampling the parameters $\btheta$, $\bbeta$ and $\balpha$ from their marginal posterior distributions and carrying out subsequent predictions. Direct computations usually entail inverting and multiplying dense matrices and also computing determinants. In software development, care is needed to avoid redundant operations and ensure numerical stability. Therefore, in the subsequent sections we describe how we use Cholesky factorizations, solve triangular systems, and minimize expensive matrix operations (e.g., dense matrix multiplications) to perform all the computations.  

\subsection{Sampling the process parameters}\label{Sec: Sampling_Process_Parameters_Generic}
Sampling from (\ref{Eq: Bayesian_Spatial_Gaussian_Generic}) employs MCMC methods, in particular Gibbs sampling and random walk Metropolis steps (e.g., \citealt{robert2010}). For faster convergence, we integrate out $\bbeta$ and $\balpha$ from the model and first sample from $\displaystyle p(\btheta\given \by) \propto p(\btheta) \times N(\by\given \bX\bmu_{\beta}, \bSigma_{y\given\theta})$,
where $\bSigma_{y\given\theta} = \bX\bSigma_{\beta}\bX' + \bZ(\btheta)\bK(\btheta)\bZ(\btheta)' + \bD(\btheta)$. This matrix needs to be constructed for every update of $\btheta$. Usually $\bD(\btheta)$ is diagonal and $\bX\bSigma_{\beta}\bX'$ is fixed, so the computation involves the matrix $\bZ(\btheta)\bK(\btheta)\bZ(\btheta)'$. Assuming that $\bZ(\btheta)$ and $\bK(\btheta)$ are computationally inexpensive to construct for each $\btheta$,  $\bZ(\btheta)\bK(\btheta)\bZ(\btheta)'$ requires $rn^2$ flops (floating point operations).  

We adopt a random-walk Metropolis step with a multivariate normal proposal (same dimension as there are parameters in $\btheta$) after transforming parameters to have support over the entire real line. This involves evaluating
\begin{equation}\label{Eq:target_theta_generic}
\log p(\btheta\given\by) = \mbox{const} + \log p(\btheta) - \frac{1}{2} \log |\bSigma_{y\given\theta}| - \frac{1}{2}Q(\btheta)\; ,
\end{equation}
where $Q(\btheta) = (\by - \bX\bmu_{\beta})'\bSigma^{-1}_{y\given \theta}(\by-\bX\bmu_{\beta})$. Generally, we compute $\bL=\mbox{\texttt{chol}}(\bSigma_{y\given \theta})$, where $\chol(\bSigma_{y\given \theta})$ returns the lower-triangular Cholesky factor $\bL$ of $\bSigma_{y\given\btheta}$. This involves $O(n^3/3)$ flops. Next, we obtain $\bu = \trsolve(\bL, \by - \bX\bmu_{\beta})$, which solves the triangular system $\bL\bu = \by - \bX\bmu_{\beta}$. This involves $O(n^2)$ flops and $Q(\btheta) = \bu'\bu$ requires another $2n$ flops. The log-determinant in (\ref{Eq:target_theta_generic}) is evaluated as $2\sum_{i=1}^n \log l_{ii}$, where $l_{ii}$ are the diagonal entries in $\bL$. Since $\bL$ has already been obtained, the log-determinant requires another $n$ steps. Therefore, the Cholesky factorization dominates the work and computing (\ref{Eq:target_theta_generic}) is achieved in $O(n^3)$ flops.

If $\bbeta$ is flat, i.e. $\bSigma_{\beta}^{-1}=\bO$, the analogue of (\ref{Eq:target_theta_generic}) is 
\begin{equation}\label{Eq:target_theta_generic_beta_flat}
\log p(\btheta\given\by) = \mbox{constant} + \log p(\btheta) - \frac{1}{2} \log |\bX'\bSigma_{y\given\beta,\theta}\bX| - \frac{1}{2} \log |\bSigma_{y\given\beta,\theta}| - \frac{1}{2}Q(\btheta),
\end{equation}
where $\bSigma_{y\given\beta,\theta} = \bZ(\btheta)\bK(\btheta)\bZ(\btheta)' + \bD(\btheta)$ and $Q(\btheta) = \by'\bSigma_{y\given\beta,\theta}^{-1}\by - \bb'(\bX'\bSigma_{y\given\beta,\theta}^{-1}\bX)^{-1}\bb$ and $\bb = \bX'\bSigma_{y\given\beta,\theta}^{-1}\by$. Computations proceed similar to the above. We first evaluate  $\bL = \mbox{\texttt{chol}}(\bSigma_{y\given \beta,\theta})$ and then obtain $[\bv:\bU] = \trsolve(\bL, [\by:\bX])$, so $\bL\bv=\by$ and $\bL\bU = \bX$. Next, we evaluate $\bW = \mbox{\texttt{chol}}(\bU'\bU)$, $\bb = \bU'\bv$ and solve $\tilde{\bb}=\trsolve(\bW,\bb)$. Finally, (\ref{Eq:target_theta_generic_beta_flat}) is evaluated as
 \begin{equation*}
    \log p(\btheta) - \sum^p_{i=1}\log w_{i,i} - \sum^n_{i=1} l_{i,i} - \frac{1}{2}(\bv'\bv - \tilde{\bb}'\tilde{\bb}),
  \end{equation*}
where $w_{i,i}$'s and $l_{ii}$'s are the diagonal elements in $\bW$ and $\bL$ respectively. The number of flops is again of cubic order in $n$.

Importantly, our strategy above avoids computing inverses. We use Cholesky factorizations and solve only triangular systems. If $n$ is not large, say $\sim$$10^2$, this strategy is feasible. The use of efficient numerical linear algebra routines fetch substantial reduction in computing time (see Section~\ref{Sec: compEnv}). Our implementation employs matrix-vector multiplication and avoids dense matrix-matrix multiplications wherever possible. Multiplications involving diagonal matrices are programmed using closed form expressions and inverses are obtained by solving triangular linear systems after obtaining a Cholesky decomposition. However, when $n\sim 10^3$ or higher, the computation becomes too onerous for practical use and alternative updating strategies are required. We address this in Section~\ref{Sec: Low_Rank_Generic}

\subsection{Sampling the slope and the random effects}\label{Sec: Sampling_Slope_AND_Random_Effects}
Once we have obtained marginal posterior samples $\btheta$ from $p(\btheta\given \by)$, we can draw posterior samples of $\bbeta$ and $\balpha$ using \emph{composition sampling}. Suppose $\{\btheta^{(1)},\btheta^{(2)},\ldots,\btheta^{(M)}\}$ are $M$ samples from $p(\btheta\given\by)$. Drawing $\bbeta^{(k)} \sim p(\bbeta\given \btheta^{(k)}, \by)$ and $\balpha^{(k)}\sim p(\balpha\given \btheta^{(k)},\by)$ for $k=1,2,\ldots M$ results in $M$ samples from $p(\bbeta\given\by)$ and $p(\balpha\given\by)$ respectively. Only the samples of $\btheta$ obtained after convergence (i.e. \emph{post burn-in}) of the MCMC algorithm need to be stored. 

To elucidate further, note that $\bbeta\given \btheta,\by \sim N_p(\bB\bb, \bB)$ with mean $\bB\bb$ and variance-covariance matrix $\bB$, where
\begin{align}\label{Eq: Full_Conditional_beta_generic}
\bb = \bSigma_{\beta}^{-1}\bmu_{\beta} + \bX'\bSigma_{y\given \beta,\theta}^{-1}\by\; \mbox{ and }\; \bB = \left(\bSigma_{\beta}^{-1} + \bX'\bSigma_{y\given \beta, \theta}^{-1}\bX\right)^{-1}\;.
\end{align}
For each $k=1,2,\ldots,M$, we compute $\bB$ and $\bb$ at the current value $\btheta^{(k)}$ and draw $\bbeta^{(k)}\sim N_p(\bB\bb, \bB)$. This is achieved by computing $\bb=\bSigma_{\beta}^{-1}\bmu_{\beta} + \bU'\bv$, where $\bL = \chol(\bSigma_{y\given\beta,\theta^{(k)}})$ and $[\bv : \bU] =\trsolve(\bL, [\by : \bX])$. Next, we generate $p$ independent standard normal variables, collect them into $\bz$ and set
\begin{align}\label{Eq: Recover_beta_Generic}
 \bbeta^{(k)} = \trsolve\left(\bL_{B}',\trsolve(\bL_{B}, \bb)\right) + \trsolve(\bL_{B}',\bz)\; ,
\end{align}
where $\displaystyle \bL_{B} = \chol\left(\bSigma_{\beta}^{-1} + \bU'\bU\right)$. This completes the $k$-th iteration. After $M$ iterations, we obtain $\{\bbeta^{(1)}, \bbeta^{(2)},\ldots, \bbeta^{(M)}\}$, which are samples from $p(\bbeta\given\by)$. 

Mapping point or interval estimates of spatial random effects is often helpful in identifying missing regressors and/or building a better understanding of model adequacy. $\bSigma_{y\given \alpha,\theta} = \bX\bSigma_{\beta}\bX' + \bD(\btheta)$ and note that  $\balpha\given \btheta,\by \sim N(\bB\bb, \bB)$, where
\begin{align}\label{Eq: Full_Conditional_alpha_generic}
\bb = \bZ(\btheta)'\bSigma_{y\given \alpha, \theta}^{-1}(\by-\bX\bmu_{\beta})\; \mbox{ and }\; \bB = \left(\bK(\btheta)^{-1} + \bZ(\btheta)'\bSigma_{y\given \alpha, \theta}^{-1}\bZ(\btheta)\right)^{-1}\; .
\end{align}
The vector $\bb$ here is computed analogously as for $\bbeta$.  For each $k=1,2,\ldots,M$ we now evaluate $\bL = \chol(\bSigma_{y\given \alpha,\theta^{(k)}})$, $[\bv : \bU] = \trsolve(\bL, [\by-\bX\bmu_{\beta}: \bZ(\btheta^{(k)})])$ and set $\bb = \bU(\btheta^{(k)})'\bv$. For computing $\bB$, one could proceed as for $\bbeta$ but that would involve $\chol(\bK(\btheta))$, which may become numerically unstable for certain covariance functions (e.g., the Gaussian or the Mat\'ern with large $\nu$). For robust software performance we define $\bG(\btheta)^{-1} = \bZ(\btheta)'\bSigma_{y\given \alpha, \theta}^{-1}\bZ(\btheta)$ and utilize the identity \citep{henderson1981}
\[
 \left(\bK(\btheta)^{-1} + \bG(\btheta)^{-1}\right)^{-1} =  \bG(\btheta) - \bG(\btheta)\left(\bK(\btheta) + \bG(\btheta)\right)^{-1}\bG(\btheta)\;  
\]
to devise a numerically stable algorithm. For each $k=1,2,\ldots,M$, we evaluate $\bL = \chol(\bK(\btheta^{(k)}) + \bG(\btheta^{(k)}))$, $\bW = \trsolve(\bL, \bG(\btheta^{(k)}))$ and $\bL_{B} = \chol(\bG(\btheta)^{(k)} - \bW'\bW)$. If $\bz$ is a $r\times 1$ vector of independent standard normal variables, then we set
$\displaystyle \balpha^{(k)} = \bL_{B}\bL_{B}'\bb + \bL_{B}\bz$. The resulting $\{\balpha^{(1)}, \balpha^{(2)},\ldots, \balpha^{(M)}\}$ are samples from $p(\balpha\given \by)$.
 
We remark that estimating the spatial effects involves Cholesky factorizations for $n\times n$ positive definite linear system. The above steps ensure numerical stability but they can become computationally prohibitive when $n$ becomes large. While some savings accrue from executing the above steps only for the post \emph{burn-in} samples, for $n$ in the order of thousands we recommend the low rank spatial models offered by \pkg{spBayes} (see Sections~\ref{Sec: Low_Rank_Generic} and \ref{Sec: ppImp}).   

\subsection{The special case of low-rank models}\label{Sec: Low_Rank_Generic}
The major computational load in estimating (\ref{Eq: Bayesian_Spatial_Gaussian_Generic}) arises from unavoidable Cholesky decompositions for dense $n\times n$ positive definite matrices. The required number of flops is of cubic order and must be executed in each iteration of the MCMC. For example, when a specific for of (\ref{Eq: Bayesian_Spatial_Gaussian_Generic}) is used to analyze a dataset comprising $n=2,000$ locations and $p=2$ predictors, each iteration requires $\sim$0.3 seconds of CPU time (see Section~\ref{Sec: Example2}). Marginalization, as described in Section~\ref{Sec: Sampling_Process_Parameters_Generic}, typically require fewer iterations to converge. But even if $10,000$ iterations are required to deliver full inferential output, the associated CPU time is $\sim$50 minutes. Clearly, large spatial datasets demand specialized models.



One strategy is to specify $\bZ(\btheta)$ with $r << n$. Such models are known as \emph{low-rank} models. Specific choices for $\bZ(\btheta)$ will be discussed later -- \pkg{spBayes} models $\bZ(\btheta)$ using the predictive process (see Section~\ref{Sec: ppImp}). To elucidate how savings accrue in low-rank models, consider the marginal Gaussian likelihood obtained by integrating out $\balpha$ from (\ref{Eq: Bayesian_Spatial_Gaussian_Generic})
\begin{align*}
 p(\btheta) \times N(\bbeta\given \bmu_{\beta}, \bSigma_{\beta})\times N(\by \given \bX\bbeta, \bSigma_{y\given \beta, \theta})\; ,
\end{align*}
where $\bSigma_{y\given\theta}=\bZ(\btheta)\bK(\btheta)\bZ(\btheta)' + \bD(\btheta)$. We could have integrated out $\bbeta$ too, as in Section~\ref{Sec: Sampling_Process_Parameters_Generic}, but there is apparently no practical advantage to that. For the low-rank model, each iteration of the Gibbs sampler updates $\bbeta$ and $\btheta$ from their full conditional distributions. 

The $\bbeta$ is drawn from $N(\bB\bb, \bB)$, where $\bb$ and $\bB$ are as in (\ref{Eq: Full_Conditional_beta_generic}). 
The strategy in Section~\ref{Sec: Sampling_Slope_AND_Random_Effects} would be expensive for large $n$ because computing $\bB$, though itself $p\times p$, involves a Cholesky factorization of the $n\times n$ matrix $\bSigma_{y\given\beta,\theta}$ for every new update of $\btheta$. Instead, we utilize the Sherman-Woodbury-Morrison formula
\begin{align}\label{Eq: SWM}
 \bSigma_{y\given\beta,\theta}^{-1} &= \bD(\btheta)^{-1} - \bD(\btheta)^{-1}\bZ(\btheta)\left(\bK(\btheta)^{-1} + \bZ(\btheta)'\bD(\btheta)^{-1}\bZ(\btheta)\right)\bZ(\btheta)'\bD(\btheta)^{-1} \nonumber \\
  &= \bD(\btheta)^{-1/2}\left(\bI - \bH'\bH\right)\bD(\btheta)^{-1/2}\; ,
\end{align}
where $\bH = \trsolve(\bL, \bW')$, $\bW = \bD(\btheta)^{-1/2}\bZ(\btheta)$ and $\bL = \chol(\bK(\btheta)^{-1} + \bW'\bW)$. Next, we compute $[\bv : \bV] = \bD^{-1/2}[\by : \bX]$, $\tilde{\bV} = \bH\bV$ and set  
\begin{align}\label{Eq: Update_beta_Low_Rank_Generic}
 \bb = \bSigma_{\beta}^{-1}\bmu_{\beta} + \bV'\by - \tilde{\bV}'\bH\bv\; \mbox{ and }\; \bL_{B}=\chol(\bSigma_{\beta}^{-1} + \bV'\bV - \tilde{\bV}'\tilde{\bV})\; .
\end{align}
We perform the above operations for each iteration in the Gibbs sampler, using the current update of $\btheta$, and sample the $\bbeta$ as in (\ref{Eq: Recover_beta_Generic}).  

We update process parameters $\btheta$ using a random-walk Metropolis step with target log-density
\begin{align}\label{Eq: Update_theta_Low_Rank_Generic}
 \log p(\btheta\given \by) = \mbox{const.} + \log p(\btheta) - \frac{1}{2}\log |\bSigma_{y\given\beta,\theta}| - \frac{1}{2} Q(\btheta)\; ,
\end{align}
where $Q(\btheta) = (\by-\bX\bbeta)'\bSigma_{y\given \beta,\theta}^{-1}(\by-\bX\bbeta)$. Having obtained $\bH$ as above, we evaluate $\bv = \bD(\btheta)^{-1/2}(\by - \bX\bbeta)$, $\bw=\bH\bv$, $\bT = \chol(\bI_r - \bH\bH')$  and compute (\ref{Eq: Update_theta_Low_Rank_Generic}) as
\begin{equation*}
    \log p(\btheta) - \frac{1}{2}\sum^n_{i=1}\log d_{i,i}(\btheta) + \sum^{n^\ast}_{i=1} \log t_{i,i} - \frac{1}{2}(\bv'\bv - \bw'\bw)\; ,
  \end{equation*}
where $d_{ii}(\btheta)$ and $t_{ii}$ are the diagonal entries of $\bD(\btheta)$ and $\bT$ respectively.

Once the Gibbs sampler has converged and we have obtained posterior samples for $\bbeta$ and $\btheta$, obtaining posterior samples for $\balpha$ can be achieved following closely the description in Section~\ref{Sec: Sampling_Slope_AND_Random_Effects}. In fact, since we the posterior samples of $\bbeta$ are already available, we can draw $\balpha$ from its full-conditional distribution, given \emph{both} $\bbeta$ and $\btheta$. This amounts to replacing $\bmu_{\beta}$ with $\bbeta$ and $\bSigma_{y\given \alpha,\theta}$ with $\bD(\btheta)$ in (\ref{Eq: Full_Conditional_alpha_generic}). The algorithm now proceeds exactly as in Section~\ref{Sec: Sampling_Slope_AND_Random_Effects} and we achieve computational savings as $\bD(\btheta)$ is usually cheaper to handle than $\bSigma_{y\given \alpha,\theta}$.

\subsection{Spatial predictions}\label{Sec: Spatial_Predictions_Kriging}  
To predict a random $t\times 1$ vector $\by_0$ associated with a $t\times p$ matrix of predictors, $\bX_0$, we assume that
\begin{align}\label{Eq: Prediction_Joint_Distribution_generic}
 \begin{bmatrix}\by\\ \by_0\end{bmatrix}\given \bbeta, \btheta \sim N_{t+n}\left(\begin{bmatrix}\bX\\ \bX_0\end{bmatrix}\bbeta,\; \begin{bmatrix} \bC_{11}(\btheta) & \bC_{12}(\btheta) \\ \bC_{12}(\btheta)' & \bC_{22}(\btheta) \end{bmatrix}\right)\; ,
\end{align}
where $\bC_{11}(\btheta) = \bSigma_{y\given\beta, \theta}$, $\bC_{12}(\btheta)$ is the $n\times t$ cross-covariance matrix between $\by$ and $\by_0$, and $\bC_{22}(\btheta)$ is the variance-covariance matrix for $\by_0$. How these are constructed is crucial for ensuring a legal probability distribution or, equivalently, a positive-definite variance-covariance matrix for $(\by',\by_0')'$ in (\ref{Eq: Prediction_Joint_Distribution_generic}). A legitimate joint distribution will supply a conditional distribution $p(\by_0\given\by,\bbeta,\btheta)$, which is normal with mean and variance
\begin{align}\label{Eq: Prediction_Conditional_Distribution_generic}
 \bmu_{p} &= \bX_0\bbeta + \bC_{12}(\btheta)'\bC_{11}(\btheta)^{-1}(\by-\bX\bbeta)\mbox{ and } \bSigma_{p} = \bC_{22}(\btheta) - \bC_{12}(\btheta)'\bC_{11}(\btheta)^{-1}\bC_{12}(\btheta)\; 
\end{align}
Bayesian prediction proceeds by sampling from the posterior predictive distribution $\displaystyle p(\by_0\given \by) = \int p(\by_0\given \by, \bbeta,\btheta)p(\bbeta,\btheta\given\by)d\bbeta d\btheta$. For each posterior sample of $\{\bbeta,\btheta\}$, we draw a corresponding $\by_0\sim N(\bmu_p,\bSigma_p)$. This produces samples from the posterior predictive distribution.

Observe that the posterior predictive computations involve only the retained MCMC samples after convergence. Furthermore, most of the ingredients to compute $\bmu_p$ and $\bSigma_p$ have already been performed while updating the model parameters. For any posterior sample $\{\bbeta^{(k)},\btheta^{(k)}\}$, we solve $[\bu : \bV] = \trsolve(\bL, [\by-\bX\bbeta^{(k)} : \bC_{12}(\btheta^{(k)})])$, where $\bL = \chol(\bC_{11}(\btheta^{(k)}))$. Next, we set $\bmu_p^{(k)} = \bX_0\bbeta^{(k)} + \bV'\bu$ and $\bSigma_p^{(k)} = \bC_{22}(\btheta^{(k)}) - \bV'\bV$ and draw $\by_0^{(k)}\sim N(\bmu_p^{(k)},\bSigma_p^{(k)})$.

Low-rank models, where $r << n$, are again cheaper here. The operations are dominated by the computation of $\bC_{12}(\btheta)'\bC_{11}(\btheta)^{-1}\bC_{12}(\btheta)$, which can be evaluated as $\bU'\bU - \bV'\bV$, where $\bU = \bD(\btheta)^{-1/2}\bC_{12}(\btheta)$, $\bV = \bH\bU$ and $\bH$ is as in (\ref{Eq: SWM}). This avoids direct evaluation of $\bC_{11}(\btheta)^{-1}$ and avoids redundant matrix operations.

Updating $\by_0^{(k)}$'s requires Cholesky factorization of $\bSigma_p$, which is $t\times t$ and can be expensive if $t$ is large. In most practical settings, it is sufficient to take $t=1$ and perform independent individual predictions. However, if the \emph{joint} predictive distribution is sought, say when full inference is desired for a function of $\by_0$, then the predictive step is significantly cheaper if we use the posterior samples of $\balpha$ as well. Now posterior predictive sampling amounts to drawing $\by_0^{(k)}\sim N(\bX_0\bbeta^{(k)} + \bZ(\btheta^{(k)})\balpha^{(k)}, \bD(\btheta^{(k)}))$, which cheap because $\bD(\btheta)$ is usually diagonal. Low rank models are especially useful here as posterior sampling for $\balpha$ is much cheaper with $r << n$.   

\section{Computing environment}\label{Sec: compEnv}
The MCMC algorithms described in the preceding sections are implemented in \pkg{spBayes} functions. These functions are written in C++ and leverage {\R}'s \emph{Foreign Language Interface} to call Fortran BLAS (Basic Linear Algebra Subprograms, see \citealt{blackford2001}) and LAPACK (Linear Algebra Package, see \citealt{anderson1999}) libraries for efficient matrix computations. Table~\ref{blas} offers a list of key BLAS and LAPACK functions used to implement the MCMC samplers. Referring to Table~\ref{blas} and following from Section~\ref{Sec: Sampling_Process_Parameters_Generic}, {\chol} corresponds to \code{dpotrf} and {\trsolve} can be either \code{dtrsv} or \code{dtrsm}, depending on the form of the equation's right-hand side. As noted previously, we try and use dense matrix-matrix multiplication, i.e., calls to \code{dgemm}, sparingly due to its computational overhead. Often careful formulation of the problem can result in fewer calls to \code{dgemm} and other \emph{expensive} BLAS level 3 and LAPACK functions.

\begin{table}[!ht]
\centering
\caption{Common BLAS and LAPACK functions used in \pkg{spBayes} function calls.} 
\begin{tabularx}{\linewidth}{ c X }
  \hline
 Function& Description\\ 
  \hline
  \code{dpotrf} &LAPACK routine to compute the Cholesky factorization of a real symmetric positive definite matrix.\\
  \code{dtrsv} & Level 2 BLAS routine to solve the systems of equations $\bA\bx = \bb$, where $\bx$ and $\bb$ are vectors and $\bA$ is a triangular matrix.\\
  \code{dtrsm} & Level 3 BLAS routine to solve the matrix equations $\bA\bX = \bB$, where $\bX$ and $\bB$ are matrices and $\bA$ is a triangular matrix.\\
  \code{dgemv} & Level 2 BLAS matrix-vector multiplication.\\
  \code{dgemm} & Level 3 BLAS matrix-matrix multiplication.\\
   \hline
\end{tabularx}\label{blas}
\end{table}

A heavy reliance on BLAS and LAPACK functions for matrix operations allows us to leverage multi-processor/core machines via threaded implementations of BLAS and LAPACK, e.g., Intel's Math Kernel Library (MKL; \url{http://software.intel.com/en-us/intel-mkl}). With the exception of \code{dtrsv}, all functions in Table~\ref{blas} are threaded in Intel's MKL. Use of MKL, or similar threaded libraries, can dramatically reduce sampler run-times. For example, the illustrative analyses offered in subsequent sections were conducted using {\R}, and hence \pkg{spBayes}, compiled with MKL on an Intel Ivy Bridge i7 quad-core processor with hyperthreading. The use of these parallel matrix operations results in a near linear speadup in the MCMC sampler's run-time with the number of CPUs---at least 4 CPUs were in use in each function call.

\pkg{spBayes} also depends on several {\R} packages including: \pkg{coda} \citep{coda} for casting the MCMC chain results as \code{coda} objects for easier posterior analysis; \pkg{abind} \citep{abind} and \pkg{magic} \citep{magic} for forming multivariate matrices, and; \pkg{Formula} \citep{formula} for interpreting symbolic model formulas.

\section{Models offered by spBayes}\label{Sec: Models_In_spBayes}
All the models offered by \pkg{spBayes} emerge as special instances of (\ref{Eq: Bayesian_Spatial_Gaussian_Generic}). The matrix $\bD(\btheta)$ is always taken to be diagonal or block-diagonal (for multivariate models). The spatial random effects $\balpha$ are assumed to arise from a partial realization of a spatial process and the spatial covariance matrix $\bK(\btheta)$ is constructed from the covariance function specifying that spatial process. To be precise, if $\{w(\bs):\bs\in \Re^d\}$ is a Gaussian spatial process with positive definite covariance function $C(\bs,\bt;\btheta)$  (see, e.g., \citealt{bochner1955}) and if $\{\bs_1,\bs_2,\ldots,\bs_r\}$ is a set of any $r$ locations in ${\cal D}$, then $\balpha = (w(\bs_1),w(\bs_2),\ldots,w(\bs_r))'$ and $\bK(\btheta)$ is its $r\times r$ covariance matrix.

\subsection{Full rank univariate Gaussian spatial regression}\label{Sec: Univariate}
For Gaussian outcomes, geostatistical models customarily regress a spatially referenced dependent variable, say $y(\bs)$, on a $p\times 1$ vector of spatially referenced predictors $\bx(\bs)$ (with an intercept) as
\begin{equation}\label{Eq: BasicUniModel}
y\left(\bs \right) = \bx\left(\bs \right)'\bbeta + w\left(\bs\right) + \varepsilon\left(\bs\right)\; ,
\end{equation}
where $\bs \in {\cal D} \subseteq \Re^2$ is a location. The residual comprises a spatial process, $w(\bs)$, and an independent white-noise process, $\varepsilon(\bs)$, that captures measurement error or micro-scale variation. With any collection of $n$ locations, say ${\cal S} = \{\bs_1,\ldots,\bs_n\}$, we assume the independent and identically distributed $\varepsilon(\bs_i)$'s follow a Normal distribution $N(0,\tau^2)$, where $\tau^2$ is called the \emph{nugget}. The $w(\bs_i)$'s provide local adjustment (with structured dependence) to the mean and capturing the effect of unmeasured or unobserved regressors with spatial pattern.

Customarily, one assumes \emph{stationarity}, which means that $C(\bs,\bt)=C(\bs-\bt)$ is a function of the separation of sites only. \emph{Isotropy} goes further and specifies $C(\bs,\bt)=C(\|\bs-\bt\|)$, where $\|\bs-\bt\|$ is the Euclidean distance between the sites $\bs$ and $\bt$. We further specify $C(\bs,\bt)=\sigma^2\rho(\bs,\bt;\bphi)$ in terms of spatial process parameters, where $\rho(\cdot;\bphi)$ is a \emph{correlation function} while $\bphi$ includes parameters quantifying rate of correlation decay and smoothness of the surface $w(\bs)$. $\mbox{Var}(w(\bs))=\sigma^2$ represents a spatial variance component. Apart from the exponential, $\rho(\bs,\bt; \bphi) = \exp(-\phi\|\bs-\bt\|)$, and the powered exponential family, $\rho(\bs,\bt; \bphi) = \exp(-\phi\|\bs-\bt\|^\alpha)$, \pkg{spBayes} also offers users the Mat\'ern correlation function
\begin{equation}\label{Eq: Matern} \rho(\|\bs-\bt\|;\bphi) = \frac{1}{2^{\nu-1}\Gamma(\nu) }(\|\bs-\bt\|\phi)^{\nu}{\cal
K}_{\nu}(\|\bs-\bt\|\phi);\;\; \phi > 0,\,\nu > 0. 
\end{equation} 
Here $\bphi=\{\phi,\nu\}$ with $\phi$ controlling the decay in spatial correlation and $\nu$ controlling process smoothness. Specifically, if $\nu$ lies between positive integers $m$ and $(m+1)$, then the spatial process $w(\bs)$ is mean-square differentiable $m$ times, but not $m+1$ times. Also, $\Gamma$ is the usual Gamma function while ${\cal K}_{\nu}$ is a modified Bessel function of the second kind with order $\nu$.

The hierarchical model built from (\ref{Eq: BasicUniModel}) emerges as a special case of (\ref{Eq: Bayesian_Spatial_Gaussian_Generic}), where $\by$ is $n\times 1$ with entries $y(\bs_i)$, $\bX$ is $n\times p$ with $\bx(\bs_i)'$ as its rows, $\balpha$ is $n\times 1$ with entries $w(\bs_i)$, $\bZ(\btheta)=\bI_n$, $\bK(\btheta)$ is $n\times n$ with entries $C(\bs_i,\bs_j;\btheta)$ and $\bD(\btheta)=\tau^2\bI_n$. We denote by $\btheta$ the set of process parameters in $\bK(\btheta)$ and $\bD(\btheta)$. Therefore, with the Mat\'ern covariance function in (\ref{Eq: Matern}), we define $\btheta = \{\sigma^2,\phi, \nu, \tau^2\}$. 

\subsubsection{Example}\label{Sec: Example1}
The marginalized specification of (\ref{Eq: BasicUniModel}) is implemented in the \code{spLM} function. The primary output of this function is posterior samples of $\btheta$. As detailed in the preceding sections, sampling is conducted using a Metropolis algorithm. Hence, users must specify Metropolis proposal variances, i.e., \emph{tuning} values, and monitor acceptance rates for these parameters. Alternately, an adaptive MCMC Metropolis-within-Gibbs algorithm, proposed by \cite{roberts2009}, is available for a more automated function call.

A key advantage of the first stage Gaussian model is that samples from the posterior distribution of $\bbeta$ and $\bw$ can be recovered in a posterior predictive fashion, given samples of $\btheta$. In practice we often choose to only use a subset of post burn-in $\btheta$ samples to collect corresponding samples of $\bbeta$ and $\bw$. This \emph{composition} sampling, detailed in Section~\ref{Sec: Sampling_Slope_AND_Random_Effects}, is conducted by passing a \code{spLM} object to the \code{spRecover} function. 

\begin{figure}[!ht]
\begin{center}
\subfigure{\includegraphics[width=6cm]{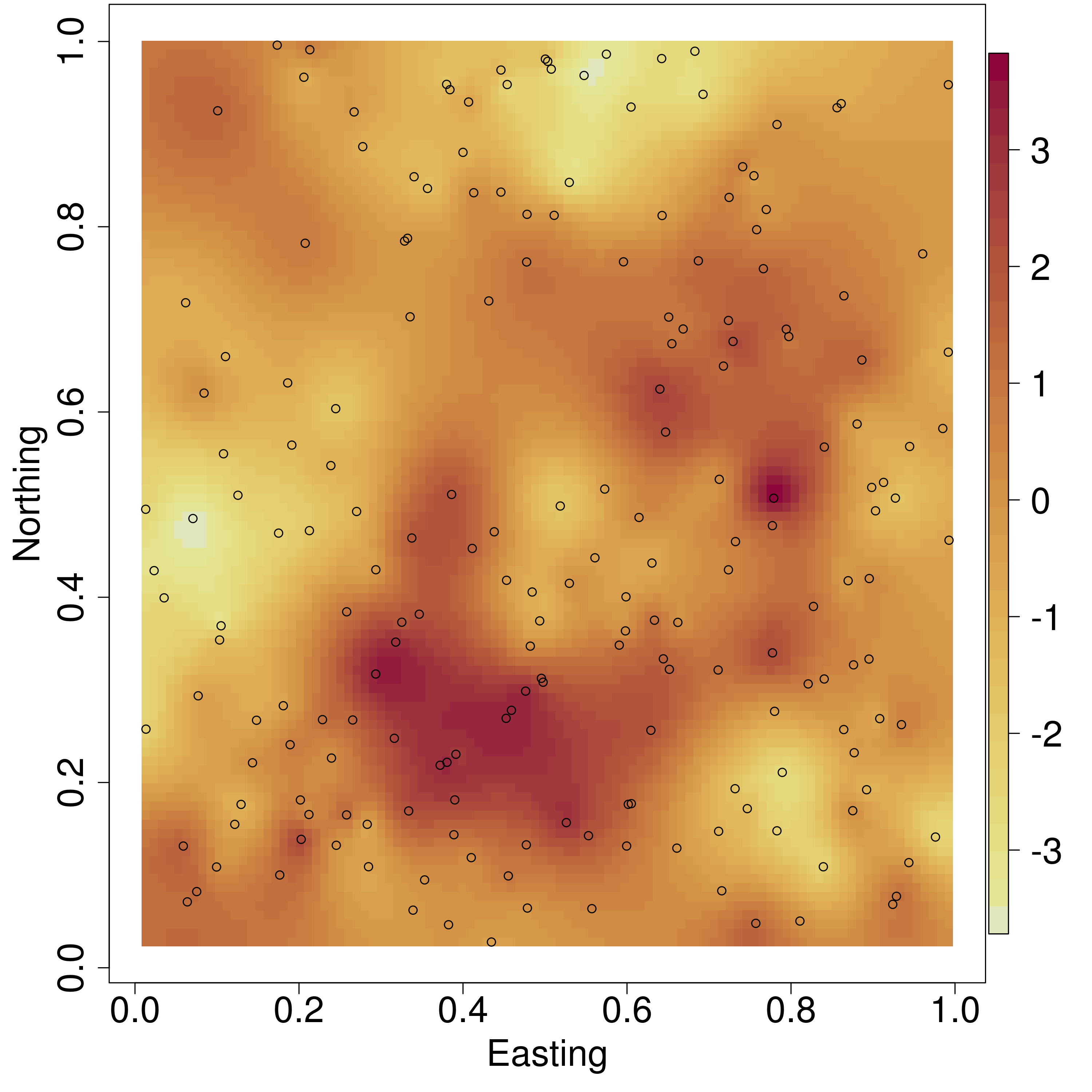}\label{fig:data1WObs}}
\subfigure{\includegraphics[width=6cm]{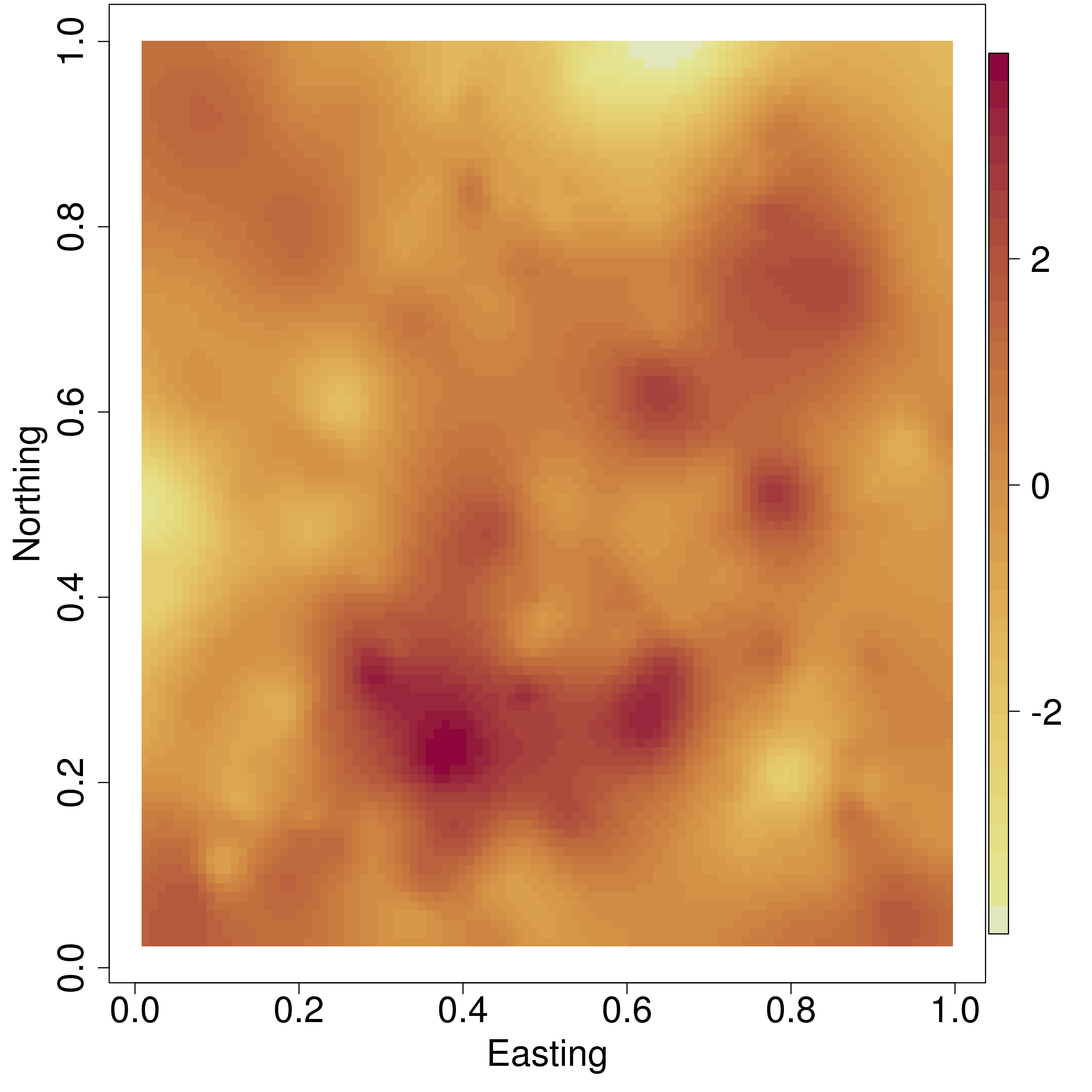}\label{fig:data1WHat}}
\end{center}
\caption{Interpolated surface of the observed \subref{fig:data1WObs} and estimated \subref{fig:data1WHat} spatial random effects.}
\label{fig:data1W}
\end{figure}

An analysis of a synthetic dataset serves to illustrate use of the \code{spLM} and \code{spRecover} functions. The data are formed by drawing $200$ observations from (\ref{Eq: BasicUniModel}) within a unit square domain. The model mean includes an intercept and covariate with associated coefficients $\beta_0=1$ and $\beta_1=5$, respectively. Model residuals are generated using an exponential spatial correlation function, with $\tau^2=1$, $\sigma^2=2$ and $\phi=6$. This choice of $\phi$ corresponds to an \emph{effective spatial range} of 0.5 distance units. For our purposes, the effective spatial range is the distance at which the correlation equals 0.05. Figure~\ref{fig:data1W} provides a surface plot of the observed spatial random effects along with the location of the $200$ observations. 

All \code{spLM} function arguments, and those of others functions highlighted in this paper, are defined in the package manual available on CRAN. Here we illustrate only some of the possible argument specifications. In addition to a symbolic model statement, the \code{spLM} function requires the user to specify: $i$) the number of MCMC samples to collect; $ii$) prior distribution, with associated hyperpriors for each parameter; $iii$) starting values for each parameter, and; $iv$) tuning values for each parameter, unless the adaptive MCMC option is chosen via the \code{amcmc} argument.

For this analysis, we assume an inverse-Gamma (IG) distribution for the variance parameters, $\tau^2$ and $\sigma^2$. These distributions are assigned \emph{shape} and \emph{scale} hyperpriors equal to 2 and 1, respectively. With a shape of 2, the mean of the IG is equal to the scale and the variance is infinite. In practice, the choice of the scale value can be guided by exploratory data analysis using a variogram or similar tools that provide estimates of the spatial and non-spatial variances. The spatial decay parameter $\phi$ is assigned a uniform (U) prior with support that covers the extent of the domain. Here, we assume $\phi$ lies in the interval between 0.1 to 1 in distance units, i.e., working from our definition of the effective spatial range this corresponds to the prior U$(-log(0.05)/1, -log(0.05)/0.1)$. In the code below, we define these priors along with the other necessary arguments that are passed to \code{spLM}. The resulting posterior samples of $\btheta$ are summarized using the \pkg{coda} package's \code{summary} function and each parameter's posterior distribution median and 95\% credible interval (CI) is printed.

\begin{Schunk}
\begin{Sinput}
R> n.samples <- 5000
R> starting <- list("tau.sq"=1, "sigma.sq"=1, "phi"=6)
R> tuning <- list("tau.sq"=0.01, "sigma.sq"=0.01, "phi"=0.1)
R> priors <- list("beta.Flat", "tau.sq.IG"=c(2, 1),
+                 "sigma.sq.IG"=c(2, 1), "phi.Unif"=c(3, 30))
R> m.i <- spLM(y~X-1, coords=coords, starting=starting,
+              tuning=tuning, priors=priors, cov.model="exponential",
+              n.samples=n.samples, n.report=2500)
\end{Sinput}
\begin{Soutput}
----------------------------------------
	General model description
----------------------------------------
Model fit with 200 observations.

Number of covariates 2 (including intercept if specified).

Using the exponential spatial correlation model.

Number of MCMC samples 5000.

Priors and hyperpriors:
	beta flat.
	sigma.sq IG hyperpriors shape=2.00000 and scale=1.00000
	tau.sq IG hyperpriors shape=2.00000 and scale=1.00000
	phi Unif hyperpriors a=3.00000 and b=30.00000
-------------------------------------------------
		Sampling
-------------------------------------------------
Sampled: 2500 of 5000, 50.00%
Report interval Metrop. Acceptance rate: 66.12%
Overall Metrop. Acceptance rate: 66.12%
-------------------------------------------------
Sampled: 5000 of 5000, 100.00%
Report interval Metrop. Acceptance rate: 64.80%
Overall Metrop. Acceptance rate: 65.46%
-------------------------------------------------
\end{Soutput}
\begin{Sinput}
R> burn.in <- floor(0.75*n.samples)
R> round(summary(window(m.i$p.theta.samples, 
+                       start=burn.in))$quantiles[,c(3,1,5)],2)
\end{Sinput}
\begin{Soutput}
          50% 2.5% 97.5%
sigma.sq 2.66 1.56  6.78
tau.sq   0.85 0.43  1.28
phi      7.17 3.01 14.94
\end{Soutput}
\end{Schunk}

Samples from the posterior distribution of $\bbeta$ and $\bw$ are then obtained by calling the \code{spRecover} function as illustrates in the code below. The samples are again returned as a \pkg{code} object that can be summarized accordingly. 

\begin{Schunk}
\begin{Sinput}
R> m.i <- spRecover(m.i, start=burn.in, thin=5, n.report=100)
\end{Sinput}
\begin{Soutput}
-------------------------------------------------
		Recovering beta and w
-------------------------------------------------
Sampled: 99 of 251, 39.44%
Sampled: 199 of 251, 79.28%
\end{Soutput}
\begin{Sinput}
R> round(summary(m.i$p.beta.recover.samples)$quantiles[,c(3,1,5)],2)
\end{Sinput}
\begin{Soutput}
    50%  2.5% 97.5%
X1 0.71 -0.78  1.77
X2 4.96  4.79  5.17
\end{Soutput}
\end{Schunk}

In practice, it is often useful to pass the mean or median of each location's spatial random effect distribution through an interpolator to generate a surface plot. These surface estimates can be created using the \code{mba.surf} function available in the \pkg{MBA} package and plotted using the \code{image} or \code{image.plot} functions from the \pkg{graphics} and \pkg{fields} packages, respectively. Such a surface is presented in Figure~\ref{fig:data1WHat} and matches closely the one depicting the synthetic data random effects in Figure~\ref{fig:data1WObs}.

\begin{Schunk}
\begin{Sinput}
R> w.hat <- apply(m.i$p.w.recover.samples, 1, median)
R> w.hat.surf <- mba.surf(cbind(coords, w.hat), 
+                         no.X=res, no.Y=res, extend=TRUE)$xyz.est
R> par(mar=c(5,5,5,5))
R> image.plot(w.hat.surf, xlab="Easting", ylab="Northing", xaxs="r", yaxs="r", 
+             cex.lab=2, cex.axis=2)
\end{Sinput}
\end{Schunk}
\subfigcapmargin = .5cm

As discussed in Section~\ref{intro}, reducing computing time was a key objective in reformulating and rewriting functions in \pkg{spBayes}. This same analysis conduced using the previous implementation of \code{spLM}, in version 0.2-4, required $\sim$8 minutes to generate $5,000$ MCMC samples of $\btheta$. The previous implementation updated $\bbeta$ from its full conditional distribution in each MCMC iteration and sampled $\btheta$ using a Metropolis algorithm that did not take advantage of triangular solvers and other efficient computational approaches detailed in the preceding sections. For comparison, the current version of \code{spLM} generates the same number of samples in 0.069 minutes.

\subsection{Low-rank predictive process models}\label{Sec: ppImp}    
\pkg{spBayes} offers low-rank models that allow the user to choose and fix $r << n$ within a hierarchical linear mixed model framework such as (\ref{Eq: Bayesian_Spatial_Gaussian_Generic}). Given the same modeling scenario as in Section~\ref{Sec: Univariate}, the user chooses $r$ locations, say ${\cal S}^* = \{\bs_1^*, \bs_2^*,\ldots,\bs_r^*\}$, and defines the process 
\begin{equation}\label{Eq: Predictive_Process}
\tildew(\bs) = \mbox{E}[w(\bs)\given w(\bs_i^*),i=1,2,\ldots,r]\; . 
\end{equation}
\cite{banerjee2008} call $\tildew(\bs)$ the \emph{predictive process}. Replacing $w(\bs)$ with $\tildew(\bs)$ in (\ref{Eq: BasicUniModel}) yields the predictive process counterpart of the univariate Gaussian spatial regression model.

The predictive process produces a low-rank model and can be cast into (\ref{Eq: Bayesian_Spatial_Gaussian_Generic}). For example, if we take $\balpha$ to the $r\times 1$ random vector with $w(\bs_i^*)$ as its entries, then the predictive process counterpart of (\ref{Eq: BasicUniModel}) is obtained from (\ref{Eq: Bayesian_Spatial_Gaussian_Generic}) with $\bD(\btheta)=\tau^2\bI$, $\bK(\btheta)=\bC^*(\btheta)$ and $\bZ(\btheta) = {\cal C}(\btheta)'\bC^{*}(\btheta)^{-1}$, where ${\cal C}(\btheta)'$ is $n\times r$ whose entries are the covariances between $w(\bs_i)$'s and $w(\bs_j^*)$'s and $\bC^*(\btheta)^{-1}$ is the $r\times r$ covariance matrix of the $w(\bs_i^*)$'s.

When employing the computational strategy for generic low-rank models described in Section~\ref{Sec: Low_Rank_Generic}, an alternative, but equivalent, parametrization  is obtained by letting $\bK(\btheta)=\bC^*(\btheta)^{-1}$ and $\bZ(\btheta) = {\cal C}(\btheta)'$. This has the added benefit of avoiding the computation of $\bC^*(\btheta)^{-1}$, which, though not expensive for low-rank models, can become numerically unstable depending upon the choice of the covariance function. Now $\balpha\sim N(\bzero,\bC^*(\btheta)^{-1})$ is no longer a vector of process realizations over the knots but it still is an $r\times 1$ random vector with a legitimate probability law. If the spatial effects over the knots are desired, they can be easily obtained from the posterior samples of $\balpha$ and $\btheta$ as $\bC^*(\btheta)\balpha$.  

We also offer an improvement over the predictive process, which attempts to capture the residual from the low-rank approximation by adjusting for the residual variance (see, e.g., \citealt{finley2009}). The difference between the spatial covariance matrices for the full rank model (\ref{Eq: BasicUniModel}) and the low-rank model is $\bC_w(\btheta) - \bZ(\btheta)\bK(\btheta)\bZ(\btheta)'$, where $\bC_w(\btheta)$ is the $n\times n$ covariance matrix of the spatial random effects for (\ref{Eq: BasicUniModel}). 

The modified predictive process model approximates this ``residual'' covariance matrix by absorbing its diagonal elements into $\bD(\btheta)$. Therefore, $\bD(\btheta) = \mbox{diag}\{\bC_w(\btheta) - \bZ(\btheta)\bK(\btheta)\bZ(\btheta)'\} + \tau^2\bI_n$, where $\mbox{diag}(\bA)$ denotes the diagonal matrix formed with the diagonal entries of $\bA$.  The remaining specifications for $\bZ(\btheta)$, $\bK(\btheta)$ and $\balpha$ in (\ref{Eq: Bayesian_Spatial_Gaussian_Generic}) remain the same as for the predictive process. 

We often refer to the modified predictive process as $\tildew_{\varepsilon}(\bs) = \tildew(\bs) + \tilde{\epsilon}(\bs)$, where $\tildew(\bs)$ is the predictive process and $\tilde{\epsilon}(\bs)$ is an independent process with zero mean and variance given by $\mbox{var}\{w(\bs)\}-\mbox{var}\{\tildew(\bs)\}$. In terms of the covariance function of $w(\bs)$, the variance of $\tilde{\epsilon}(\bs)$ is $C(\bs,\bs;\btheta)-\bc(\bs,\btheta)'\bC^*(\btheta)^{-1}\bc(\bs)$, where $\bc(\bs)$ is the $r\times 1$ vector of covariances between $w(\bs)$ and $w(\bs^*j)$ as its entries. Also, $\bw^*$, $\tildebw$ and $\tildebw_{\epsilon}$ denote the collection of $w(\bs_i^*)$'s over the $r$ knots, $\tildew(\bs_i)$'s over the $n$ locations and $\tildew_{\epsilon}(\bs_i)$'s over the $n$ locations respectively.

A key issue in low-rank models is the choice of knots. Given a computationally feasible $r$ one could fix the knot location using grid over the extent of the domain, space-covering design (e.g., \citealt{royle1998}), or more sophisticated approach aimed at minimizing a predictive variance criterion (see, e.g., \citealt{finley2009}; \citealt{guhaniyogi2011}). In practice, if the observed locations are evenly distributed across the domain, we have found relatively small difference in inference based on knot locations chosen using a grid, space-covering design, or other criterion. Rather, it is the number of knots locations that has the greater impact on parameter estimates and subsequent prediction. Therefore, we often investigate sensitivity of inference to different knot intensities, within a computationally feasible range.

\subsubsection{Example}\label{Sec: Example2}
Moving from (\ref{Eq: BasicUniModel}) to its predictive process counterpart is as simple as passing a $r\times 2$ matrix of knot locations, via the \code{knots} argument, to the \pkg{spLM} function. Choice between the non-modified and modified predictive process model, i.e., $\tildew(\bs)$ and $\tildew_{\varepsilon}(\bs)$, is specified using the \code{modified.pp} logical argument. Passing a \pkg{spLM} object, specified for a predictive process model, to \code{spRecover} will yield posterior samples from $\tildebw$ or $\tildebw_{\varepsilon}$ and $\bw^{*}$.

We construct a second synthetic dataset using the same model and parameter values from Section~\ref{Sec: Example1}, but now generate $2,000$ observations. Parameters are then estimated using the following candidate models: $i$) non-modified predictive process with 25 knot grid; $ii$) modified predictive process with 25 knot grid; $iii$) non-modified predictive process with 100 knot grid, and; $iv$) modified predictive process with 100 knot grid. 

\begin{figure}[!ht]
\begin{center}
\subfigure{\includegraphics[width=6cm]{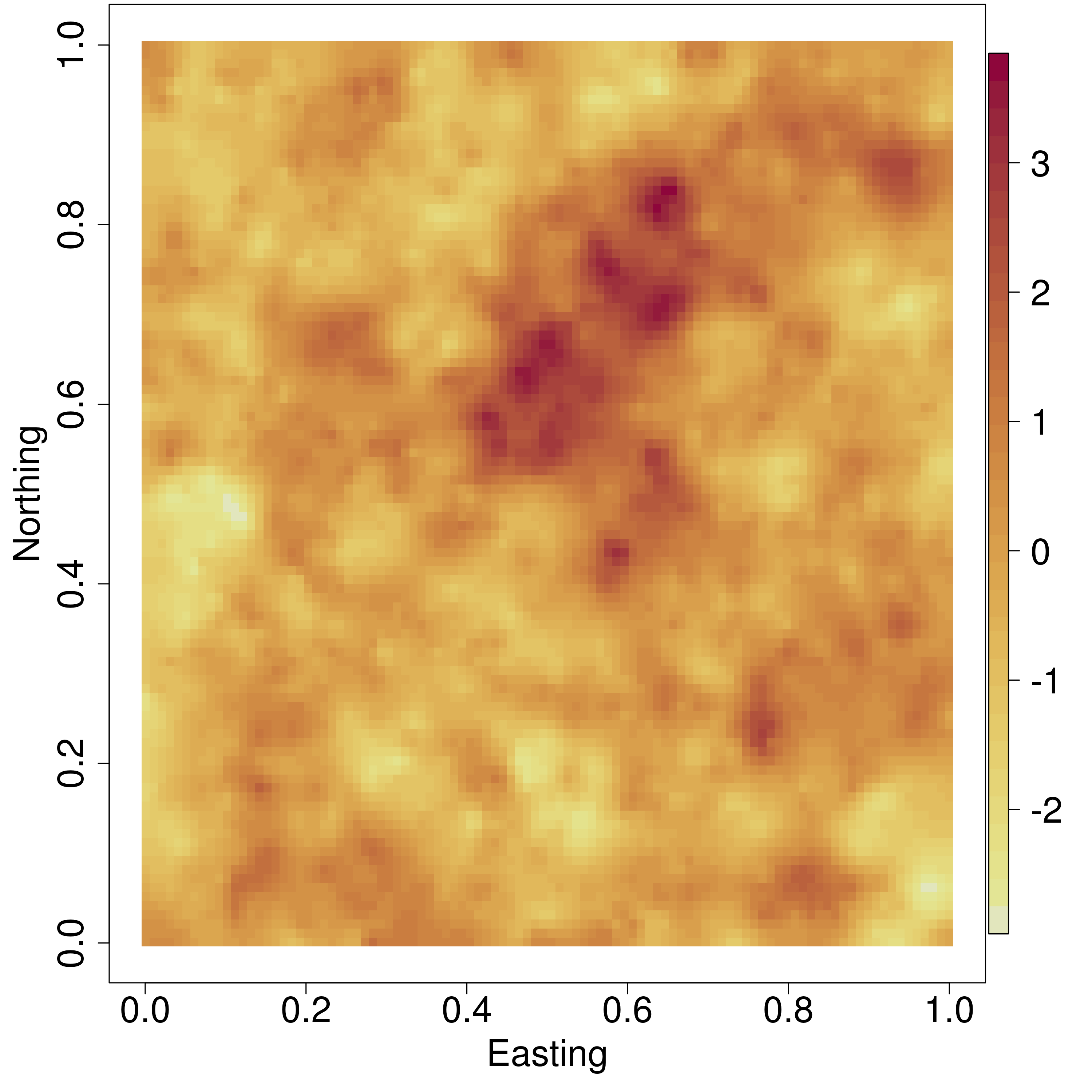}\label{fig:data2W}}
\subfigure{\includegraphics[width=6cm]{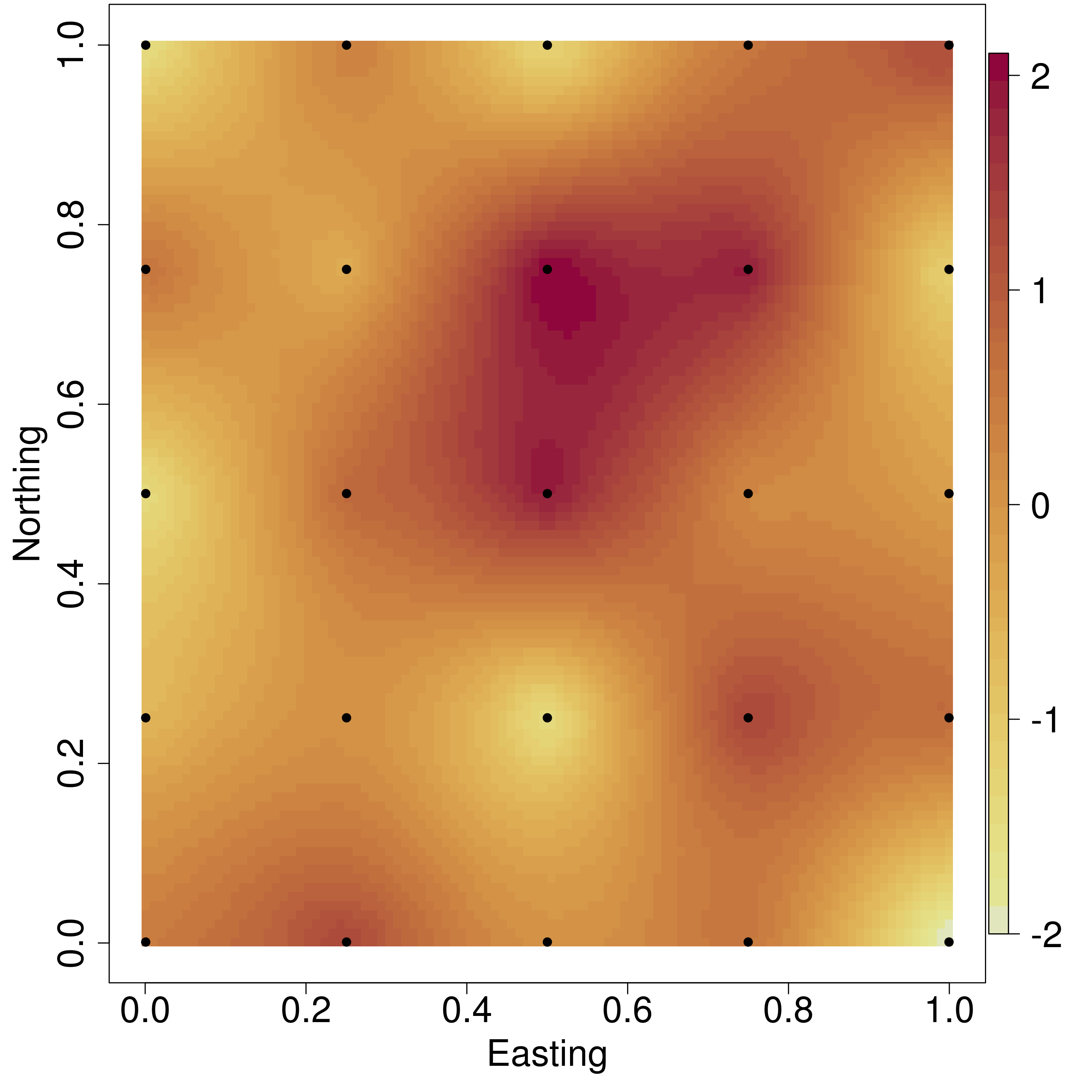}\label{fig:wSurfI}}
\subfigure{\includegraphics[width=6cm]{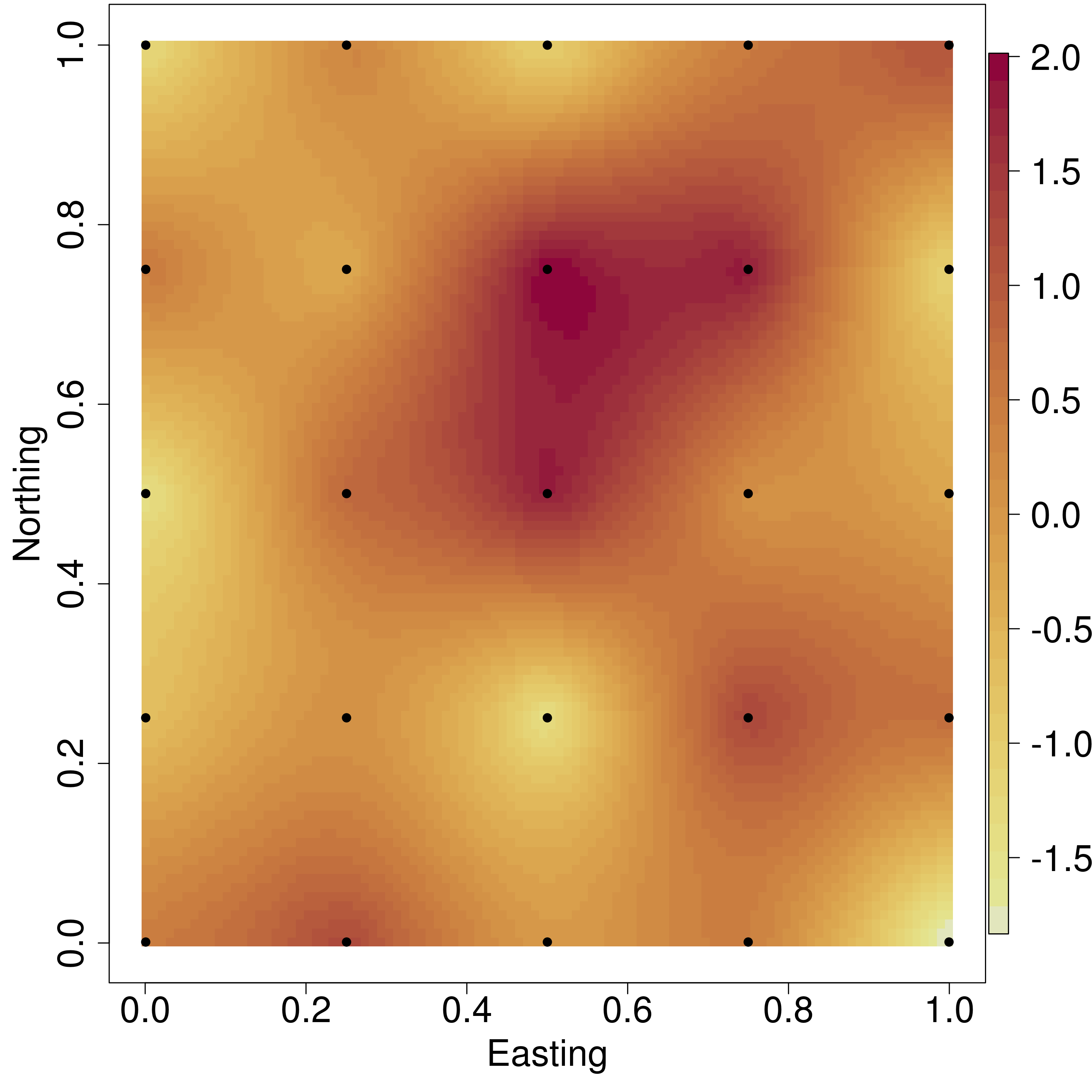}\label{fig:wSurfII}}
\subfigure{\includegraphics[width=6cm]{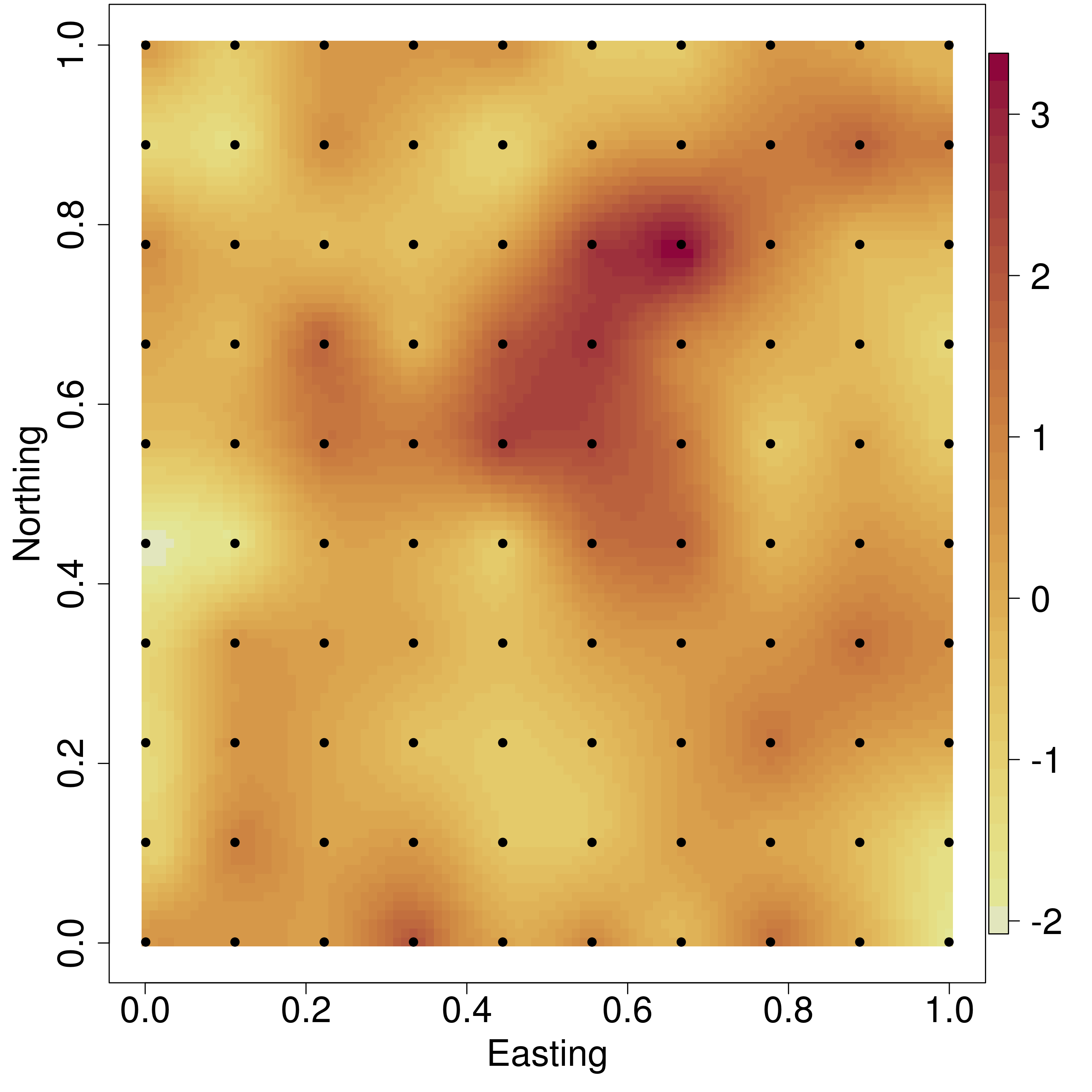}\label{fig:wSurfIII}}
\subfigure{\includegraphics[width=6cm]{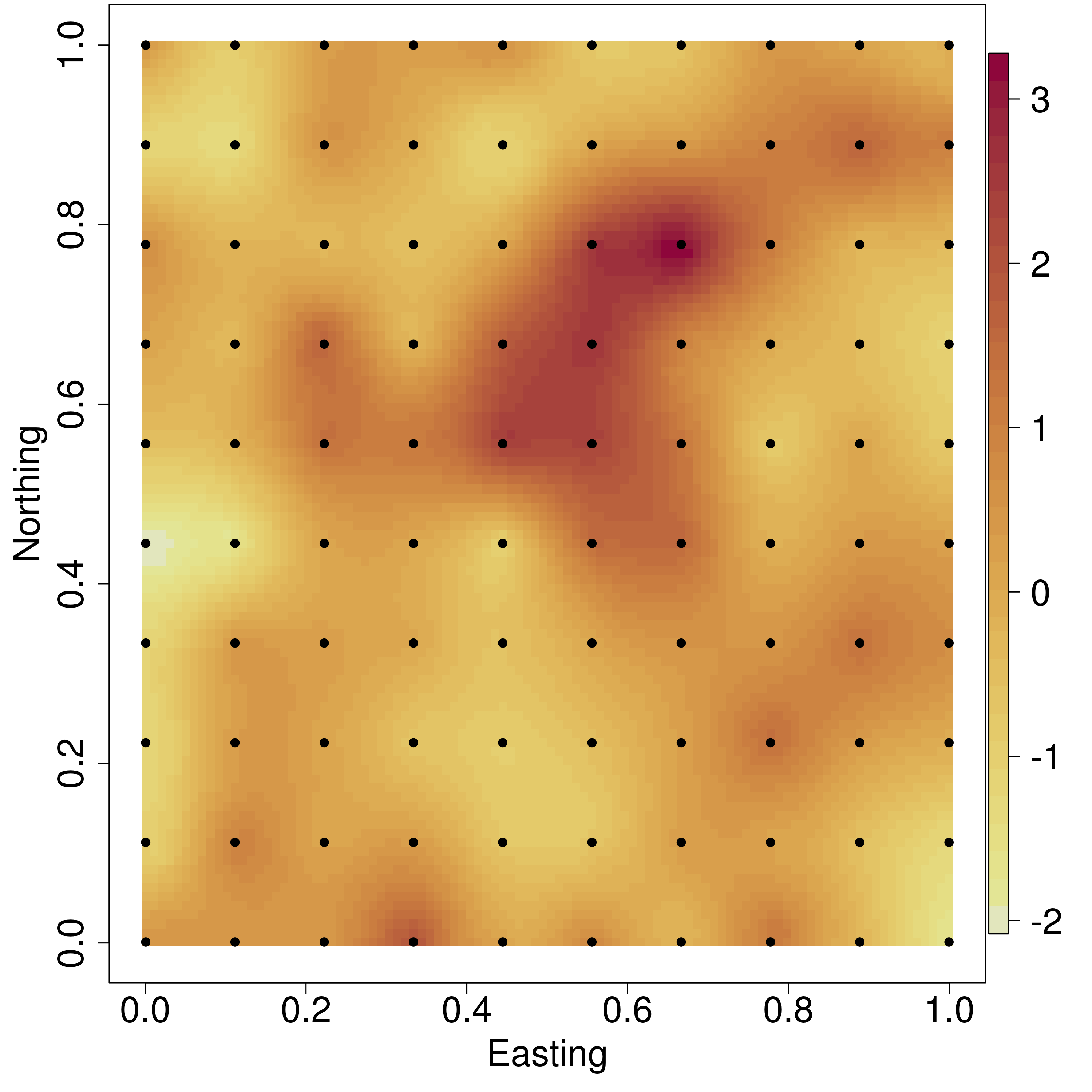}\label{fig:wSurfIV}}
\subfigure{\includegraphics[width=6cm]{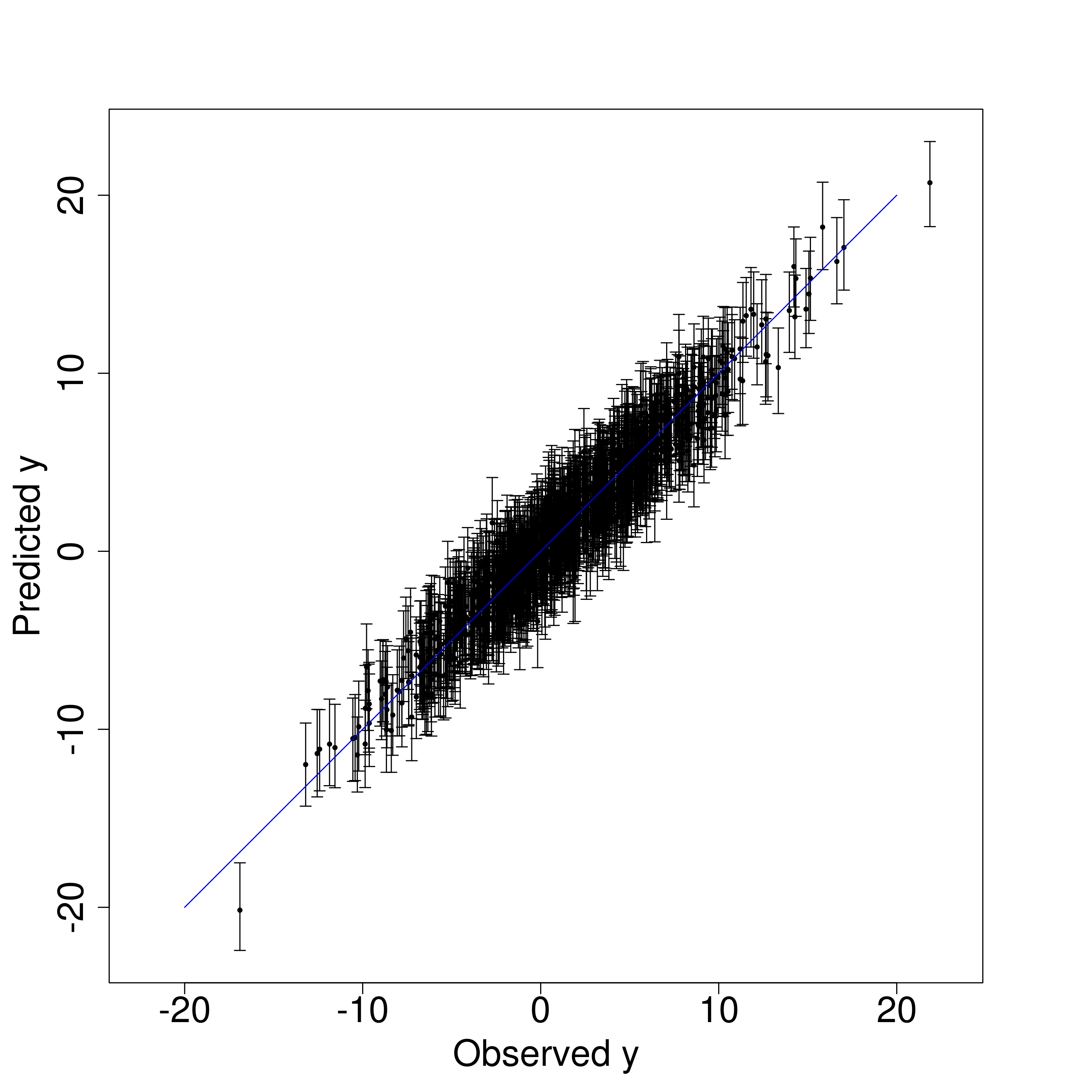}\label{fig:yVsYHat}}
\end{center}
\caption{Interpolated surfaces of the \subref{fig:data2W} observed spatial random effects and \subref{fig:wSurfI}, \subref{fig:wSurfII}, \subref{fig:wSurfIII}, \subref{fig:wSurfIV} are the estiamted spatial random effects from models $i$, $ii$, $iii$, and $iv$, respectively. Filled circile symbols in \subref{fig:wSurfI}, \subref{fig:wSurfII}, \subref{fig:wSurfIII}, \subref{fig:wSurfIV} show the location of predictive process knots. \subref{fig:yVsYHat} plots holdout observed versus candidate model $iv$ predicted median and 95\% CI intervals with 1:1 line.}
\label{fig:data2W}
\end{figure}

The \code{spLM} call for the 25 knot non-modified predictive process model is given below. The \code{starting}, \code{priors}, and \code{tuning} arguments are taken from Section~\ref{Sec: Example1}. As noted above, the \code{knots} argument invokes the predictive process model. The value portion of this argument \code{c(5, 5, 0)} specifies a 5 by 5 knot grid with should be placed over the extent of the observed locations. The third value in this vector controls the extent of the this grid, e.g., one may want the knot grid to extend beyond the convex haul of the observed locations. The placement of these knots is illustrated in Figure~\ref{fig:wSurfI}. Users can also pass in their own knot locations via the \code{knots} argument.

\begin{Schunk}
\begin{Sinput}
R> m.i <- spLM(y~X-1, coords=coords, knots=c(5, 5, 0), starting=starting,
+              tuning=tuning, priors=priors, cov.model="exponential",
+              modified.pp=FALSE, n.samples=n.samples, n.report=2500)
\end{Sinput}
\begin{Soutput}
----------------------------------------
	General model description
----------------------------------------
Model fit with 2000 observations.

Number of covariates 2 (including intercept if specified).

Using the exponential spatial correlation model.

Using non-modified predictive process with 25 knots.

Number of MCMC samples 5000.

Priors and hyperpriors:
	beta flat.
	sigma.sq IG hyperpriors shape=2.00000 and scale=1.00000
	tau.sq IG hyperpriors shape=2.00000 and scale=1.00000
	phi Unif hyperpriors a=3.00000 and b=30.00000
-------------------------------------------------
		Sampling
-------------------------------------------------
Sampled: 2500 of 5000, 50.00%
Report interval Metrop. Acceptance rate: 35.20%
Overall Metrop. Acceptance rate: 35.20%
-------------------------------------------------
Sampled: 5000 of 5000, 100.00%
Report interval Metrop. Acceptance rate: 32.24%
Overall Metrop. Acceptance rate: 33.72%
-------------------------------------------------
\end{Soutput}
\end{Schunk}

\begin{table}[!ht]
\centering
\caption{Parameter estimates and run-time (wall time) in minutes for candidate predictive process models. Parameter posterior summary 50 (2.5, 97.5) percentiles.} 
\label{tab1}
\begin{tabular}{cccccc}
  \hline
 & true & i & ii & iii & iv \\ 
  \hline
$\beta_0$ & 1 & 0.64 (-0.52, 1.83) & 0.63 (-0.37, 1.62) & 0.77 (0.07, 1.4) & 0.78 (0.03, 1.48) \\ 
  $\beta_1$ & 5 & 4.99 (4.94, 5.05) & 4.99 (4.94, 5.05) & 4.98 (4.93, 5.03) & 4.98 (4.93, 5.03) \\ 
  $\sigma^2$ & 2 & 2.3 (1.45, 3.48) & 1.57 (1.04, 2.13) & 1.89 (1.19, 2.6) & 1.65 (1.23, 3.41) \\ 
  $\tau^2$ & 1 & 1.72 (1.6, 1.84) & 1.19 (0.98, 1.42) & 1.41 (1.33, 1.51) & 0.84 (0.56, 1.03) \\ 
  $\phi$ & 6 & 3.68 (3, 4.86) & 3.39 (3.03, 4.17) & 8.19 (5.62, 11.23) & 7.75 (3.93, 11.3) \\ 
  Time &  & 0.19 & 0.23 & 0.89 & 1 \\ 
   \hline
\end{tabular}
\end{table}
Table~\ref{tab1} provides parameter estimates and run-time for all candidate models. Here, the predictive process induced upward bias, described in Section~\ref{Sec: ppImp}, is seen in model $i$ and $iii$ $\tau^2$ estimates. This bias is removed by using the modified predictive process, as illustrated by model $ii$ and $iv$ variance parameter estimates. As show by the run-times, there is only a marginal difference in computation overhead between the non-modified and modified predictive process models. In most settings the modification should be used. 

For comparison with Table~\ref{tab1}, the full rank model required 5.18 minutes to generate the $5,000$ posterior samples. Also parameter estimates from the full rank model were comparable to those of model $iv$. These attractive qualities of the predictive process models do not extend to all settings. For example, if the range of spatial dependence is short relative to the spacing of the knots, then covariance parameter estimation will suffer. We are obviously forgoing some information about underlying spatial process when using an array of knots that is coarse compared to the number of observations. This is most easily seen by comparing estimated spatial random effects surfaces to the \emph{true} surface used to generate the data, as shown in Figure~\ref{fig:data2W}. This smoothing of the random effects surface can translate into diminished predictive ability and, in some cases, model parameter inference, compared to a full rank model.


\clearpage
Following from Section~\ref{Sec: Spatial_Predictions_Kriging}, given coordinates and predictors for \emph{new} locations, and a \code{spLM} object, the \code{spPredict} function returns posterior predictive samples from $\by_0$. The \code{spPredict} function provides a generic interface for prediction using most model functions in \code{spBayes}. The code below illustrates prediction using model $iv$ for $1,000$ holdout locations. Here, \code{X.ho} is the $1,000\times 2$ (i.e., $t\times p$) predictor matrix associated with the $1,000$ holdout coordinates stored in \code{coords.ho}. 

\begin{Schunk}
\begin{Sinput}
R> m.iv.pred <- spPredict(m.iv, start=burn.in, thin=2, pred.covars=X.ho, 
+                         pred.coords=coords.ho, verbose=FALSE)
\end{Sinput}
\end{Schunk}

\begin{Schunk}
\begin{Sinput}
R> y.hat <- apply(m.iv.pred$p.y.predictive.samples, 1, quants)
R> par(mar=c(5,5,5,5))
R> plot(y.ho, y.hat[1,], pch=19, cex=0.5, xlab="Observed y", ylab="Predicted y", 
+       ylim=range(y.hat), xlim=range(y.hat), cex.lab=2, cex.axis=2)
R> arrows(y.ho, y.hat[1,], y.ho, y.hat[2,], angle=90, length=0.05)
R> arrows(y.ho, y.hat[1,], y.ho, y.hat[3,], angle=90, length=0.05)
R> lines(-20:20,-20:20, col="blue")
\end{Sinput}
\end{Schunk}

Figure~\ref{fig:yVsYHat} shows the observed versus predicted values for the holdout locations. We expect the posterior predictive 95\% CIs will cover $\sim$950 of the \emph{true} values in $\by_0$. For this analysis, our coverage rate was 94.4 percent.

\section{Multivariate Gaussian spatial regression models}\label{Sec: Multivariate}
Multivariate spatial regression models consider $m$ point-referenced outcomes that are regressed, at each location, on a known set of predictors
\begin{equation}\label{Eq: Basic_Multivariate_Model}
y_j(\bs) = \bx_{j}(\bs)'\bbeta_j + w_j(\bs) + \epsilon_j(\bs)\; ,\; \mbox{ for }\; j=1,2,\ldots,m\; ,
\end{equation}
where $\bx_j(\bs)$ is a $p_j\times 1$ vector of predictors associated with outcome $j$, $\bbeta_j$ is the $p_j\times 1$ slope, $w_j(\bs)$ and $\epsilon_j(\bs)$ are the spatial and random error processes associated with outcome $y_j(\bs)$.  Customarily, we assume the unstructured residuals $\bepsilon(\bs)=(\epsilon_1(\bs),\epsilon_2(\bs),\ldots,\epsilon_m(\bs))'$ follow a zero-centered multivariate normal distribution with zero mean and an $m\times m$ dispersion matrix $\Psi$. 

Spatial variation is modeled using an $m\times 1$ Gaussian process $\bw(\bs) = (w_1(\bs),\ldots,w_m(\bs))'$, specified by a zero mean and a cross-covariance matrix $\bC_w(\bs,\bt)$ with entries being covariance between $w_i(\bs)$ and $w_j(\bt)$. \pkg{spBayes} uses the linear model of coregionalization (LMC) to specify the cross-covariance. This assumes that $\bC_w(\bs,\bt)=\bA\bM(\bs,\bt)\bA'$, where $\bA$ is $m\times m$ lower-triangular and $\bM(\bs,\bt)$ is $m\times m$ diagonal with each diagona entry a spatial correlation function endowed with its own set of process parameters. 

Suppose we have observed the $m$ outcomes in each of $b$ locations. Let $\by$ be $n\times 1$, where $n=mb$, obtained by stacking up the $\by(\bs_i)$'s over the $b$ locations. Let $\bX$ be the $n\times p$ matrix of predictors associated with $\by$, where $p=\sum_{j=1}^m p_j$, and $\bbeta$ is $p\times 1$ with the $\bbeta_j$'s stacked correspondingly. Then, the hierarchical multivariate spatial regression models arise from (\ref{Eq: Bayesian_Spatial_Gaussian_Generic}) with the following specifications: $\bD(\btheta) = \bI_b\otimes \bPsi$, $\balpha$ is $n\times 1$ formed by stacking the $\bw_i$'s and $\bK(\btheta)$ is $n\times n$ partitioned into $m\times m$ blocks given by $\bA\bM(\bs_i,\bs_j)\bA'$. The positive-definiteness of $\bK(\btheta)$ is ensured by the linear model of coregionalization \citep{gelfand2004}. \pkg{spBayes} also offers low rank multivariate models involving the predictive process and the modified predictive process that can be estimated using strategies analogous to Section~\ref{Sec: Low_Rank_Generic}. Both the full rank multivariate Gaussian model and its predictive process counterpart are implemented in the \code{spMvLM} function. Notation and additional background for fitting these models is given by \cite{banerjee2008} and \cite{finley2009} as well as example code in the \code{spMvLM} documentation examples.

\section{Non-Gaussian models}\label{Sec: Spatial_GLM}
Two typical non-Gaussian first stage settings are implemented in \pkg{spBayes}: $i$) binary response at locations modeled using logit or probit
regression, and; $ii$) count data at locations modeled using Poisson regression. \cite{diggle1998} unify the use of
generalized linear models in spatial data contexts. See also \cite{lin2000}, \cite{kammann2003} and \cite{banerjee2004}.
Here we replace the Gaussian likelihood in (\ref{Eq: Bayesian_Spatial_Gaussian_Generic}) with the assumption that $\mbox{E}[y(\bs)]$ is linear on a transformed scale, i.e., $\eta(\bs) \equiv g(\mbox{E}(y(\bs)))= \bx(\bs)'\bbeta + w(\bs)$ where $g(\cdot)$ is a suitable link function. We refer to these as spatial generalized linear models (GLM's).

With the Gaussian first stage, we can marginalize over the spatial effects and implement our MCMC over a reduced parameter space. With a binary or Poisson first stage, such marginalization is precluded and we have to update the spatial effects in running our Gibbs sampler. We offer both the traditional random-walk Metropolis as well as the adaptive random-walk Metropolis \citep{roberts2009} to update the spatial effects. \pkg{spBayes} also provides low-rank predictive process versions for spatial GLM's. The analogue of (\ref{Eq: Bayesian_Spatial_Gaussian_Generic}) is
\begin{align}\label{Eq: Bayesian_GLM_Generic}
 p(\btheta) \times N(\bbeta\given \bmu_{\beta},\bSigma_{\beta})\times N(\balpha\given\bzero,\bK(\btheta)) \times \prod_{i=1}^n f(y(\bs_i)\given \eta(\bs_i)\equiv\bx(\bs_i)'\bbeta + \bz_i(\btheta)'\balpha)\; ,
\end{align}
where $f(\cdot)$ represents a Bernoulli or Poisson density with $\eta(\bs)$ represents the mean of $y(\bs)$ on a transformed scale. This model and its predictive process counterpart is implemented in the \code{spGLM} function. These models are extended to accommodate multivariate settings, outlined in Section~\ref{Sec: Multivariate}, using the \code{spMvGLM} function.

\section{Dynamic spatio-temporal models}\label{models}
There are many different flavors of spatio-temporal data and an extensive statistical literature that addresses the most common settings. The approach adopted here applies to the setting where space is viewed as continuous, but time is assumed to be discrete. Put another way, we view the data as a time series of spatial process realizations and work in the setting of dynamic models. Building upon previous work in the setting of dynamic models by \cite{west1997}, several authors, including \cite{stroud2001} and \cite{gelfand2005}, proposed dynamic frameworks to model residual spatial and temporal dependence. These proposed frameworks are flexible and easily extended to accommodate nonstationary and multivariate outcomes.

Dynamic linear models, or state-space models, have gained tremendous popularity in recent years in fields as disparate as engineering, economics, genetics, and ecology. They offer a versatile framework for fitting several time-varying models \citep{west1997}.  \cite{gelfand2005} adapted the dynamic modeling framework to spatio-temporal models with spatially varying coefficients. Alternative adaptations of dynamic linear models to space-time data can be found in \cite{stroud2001}.

\subsection{Model specification}\label{spDynImp}
\pkg{spBayes} offers a relatively simple version of the dynamic models in \cite{gelfand2005}. Suppose, $y_t(\bs)$ denotes the observation at location $\bs$ and time $t$. We model $y_t(\bs)$ through a \emph{measurement equation} that provides a regression specification with a space-time varying intercept and  serially and spatially uncorrelated zero-centered Gaussian disturbances as measurement error $\epsilon_t(\bs)$. Next a \emph{transition equation} introduces a $p \times 1$ coefficient vector, say $\bbeta_t$, which is a purely temporal component (i.e., time-varying regression parameters), and a spatio-temporal component $u_t(\bs)$. Both these are generated through transition equations, capturing their Markovian dependence in time. While the transition equation of the purely temporal component is akin to usual state-space modeling, the spatio-temporal component is generated using Gaussian spatial processes. The overall model is written as
\begin{align}\label{Eq: Dynamic_ST_Model}
y_t(\bs) &= \bx_t(\bs)'\bbeta_t + u_t(\bs) + \epsilon_t(\bs),\quad \epsilon_t(\bs)\stackrel{ind.}\sim N(0,\tau_{t}^2)\; ;\nonumber \\
\bbeta_t &= \bbeta_{t-1} + \bet_t,\quad \bet_t\stackrel{i.i.d.}\sim
N(0,\bSigma_{\eta})\; ;\nonumber \\
u_t(\bs) &= u_{t-1}(\bs) + w_t(\bs),\quad w_t(\bs) \stackrel{ind.}{\sim} GP\left(\bzero, C_t(\cdot,\btheta_t)\right),\quad t=1,2,\ldots,N_t\;,
\end{align}
where the abbreviations $ind.$ and $i.i.d$ are \emph{independent} and \emph{independent and identically distributed}, respectively. Here $\bx_t(\bs)$ is a $p\times 1$ vector of predictors and $\bbeta_t$ is a $p\times 1$ vector of coefficients. In addition to an intercept, $\bx_t(\bs)$ can include location specific variables useful for explaining the variability in $y_t(\bs)$. The $GP(\bzero, C_t(\cdot,\btheta_t))$ denotes a spatial Gaussian process with covariance function $C_{t}(\cdot;\btheta_t)$. We customarily specify $C_{t}(\bs_1,\bs_2;\btheta_t)=\sigma_t^2\rho(\bs_1,\bs_2;\phi_t)$, where $\btheta_t = \{\sigma_t^2,\phi_t\}$ and $\rho(\cdot;\phi)$ is a \emph{correlation function} with $\phi$ controlling the correlation decay and $\sigma_t^2$ represents the spatial variance component. We further assume $\bbeta_0 \sim N(\bm_0, \bSigma_0)$ and $u_0(\bs) \equiv 0$, which completes the prior specifications leading to a well-identified Bayesian hierarchical model with reasonable dependence structures. In practice, estimation of model parameters are usually very robust to these hyper-prior specifications. Also note that (\ref{Eq: Dynamic_ST_Model}) reduces to a simple spatial regression model for $t=1$.  

We consider settings where the inferential interest lies in spatial prediction or interpolation over a region for a set of discrete time points. We also assume that the same locations are monitored for each time point resulting in a space-time matrix whose rows index the locations and columns index the time points, i.e. the $(i,j)$-th element is $y_j(\bs_i)$. Our algorithm will accommodate the situation where some cells of the space-time data matrix may have missing observations, as is common in monitoring environmental variables.

Conducting full Bayesian inference for (\ref{Eq: Dynamic_ST_Model}) is computationally onerous and \pkg{spBayes} also offers a modified predictive process counterpart of (\ref{Eq: Dynamic_ST_Model}). This is achieved by replacing $u_t(\bs)$ in (\ref{Eq: Dynamic_ST_Model}) with $\tilde{u}_t(\bs) = \sum_{k=1}^t \left[\tildew_k(\bs) + \tilde{\epsilon}_k(\bs)\right]$, where $\tildew_k(\bs)$ is the predictive process as defined in (\ref{Eq: Predictive_Process}) and the ``adjustment'' $\tilde{\epsilon}_t(\bs)$ compensates for the oversmoothing by the conditional expectation component and the consequent underestimation of spatial variability (see \citealt{finley2012}) for details.

\subsubsection{Example}
The dynamic model (\ref{Eq: Dynamic_ST_Model}) and its predictive process counterpart are implemented in the \code{spDynLM} function. Here we illustrate the full rank dynamic model using an ozone monitoring dataset that was previously analyzed by \cite{sahu2011}. This is a relatively small dataset and does not require dimension reduction. Note, however, similar to other \pkg{spBayes} models, moving from full to low-rank representation of $\bu_t$ only requires specification of knot locations via the \code{knots} argument in the model call.

\begin{figure}[!ht]
\begin{center}
\includegraphics[width=10cm]{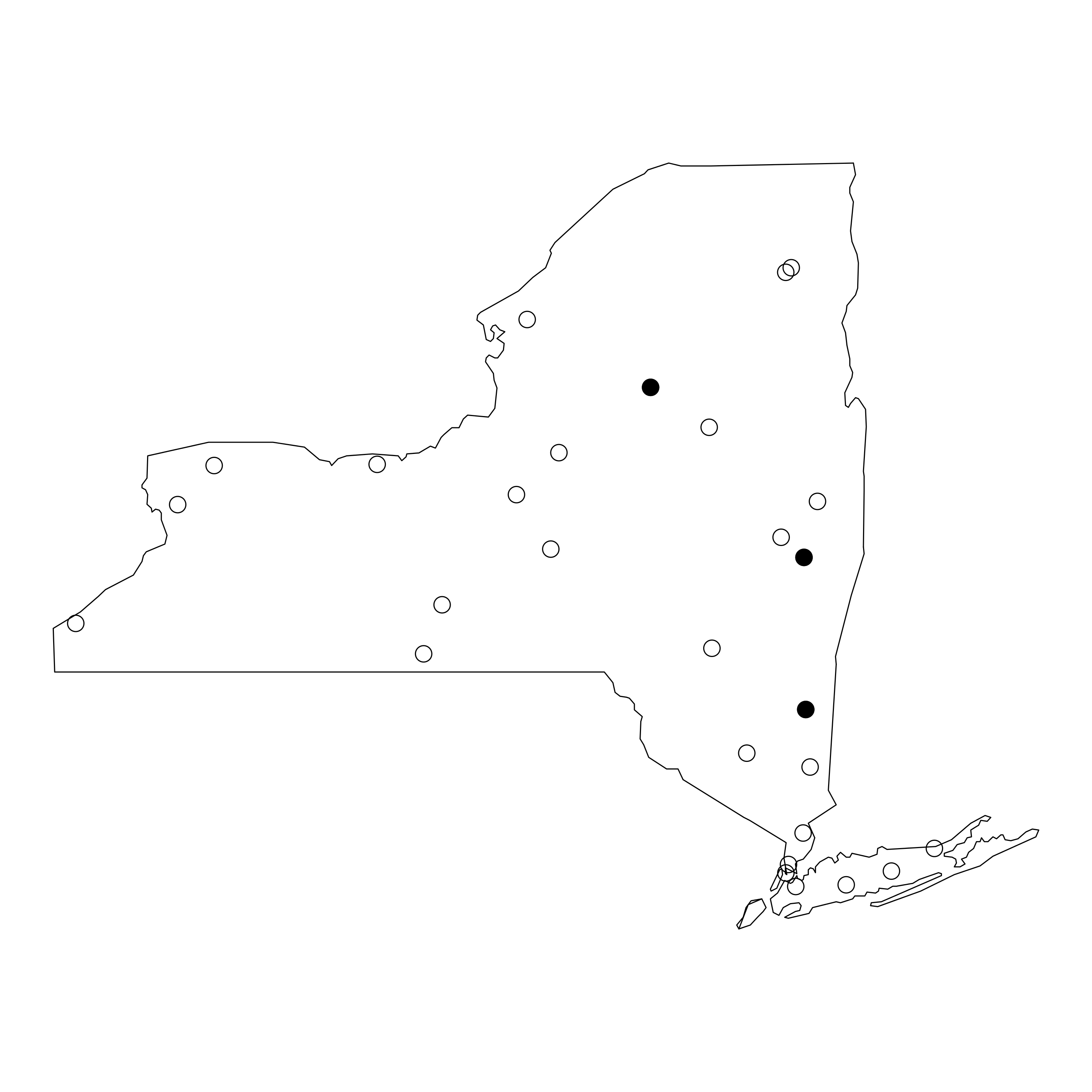}
\end{center}
\caption{Open and filled circle symbols indicate the location of 28 ozone monitoring stations across New York State. Filled circle symbols identify those stations that have half of the daily ozone measurements withheld to assess model predictive performance.}
\label{fig:ozoneStations}
\end{figure}

The dataset comprises 28 Environmental Protection Agency monitoring stations that recorded ozone from July 1 to August 31, 2006. The outcome is daily 8-hour maximum average ozone concentrations (parts per billion; O3.8HRMAX), and predictors include maximum temperature (Celsius; cMAXTMP), wind speed (knots; WDSP), and relative humidity (RM). Of the $1,736$ possible observations, i.e., $n$=28 locations times $N_t$=62 daily O3.8HRMAX measurements, $114$ are missing. In this illustrative analysis we use the predictors cMAXTMP, WDSP, and RM as well as the spatially and temporally structured residuals to predict missing O3.8HRMAX values. To gain a better sense of the dynamic model's predictive performance, we withheld half of the observations from the records of three stations for subsequent validation. Figure~\ref{fig:ozoneStations} shows the monitoring station locations and identifies those stations where data were withheld.

The first \code{spDynLM} function argument is a list of $N_t$ symbolic model statements representing the regression within each time step. This can be easily assembled using the \code{lapply} function as shown in the code below. Here too, we define the station coordinates as well as starting, tuning, and prior distributions for the model parameters. Exploratory data analysis using time step specific variograms can be helpful for defining starting values and prior support for parameters in $\btheta_t$ and $\tau^2_t$. To avoid cluttering the code, we specify the same prior for the $\phi_t$'s, $\sigma^2_t$'s, and $\tau^2_t$'s. As in the other \code{spBayes} model functions, one can choose among several popular spatial correlation functions including the exponential, spherical, Gaussian and Mat\'ern. The exponential correlation function is specified in the \code{spDynLM} call below. Unlike other model functions described in the preceding sections, the \code{spDynLM} function will accept \code{NA} $y_t(\bs)$ values. The sampler will provide posterior predictive samples for these missing values. If the \code{get.fitted} argument is \code{TRUE} then these posterior predictive samples are save along with posterior \emph{fitted} values for locations where the outcomes are observed.

\begin{Schunk}
\begin{Sinput}
R> mods <- lapply(paste("O3.8HRMAX.",1:N.t, "~cMAXTMP.",1:N.t, 
+                       "+WDSP.",1:N.t, "+RH.",1:N.t, sep=""), as.formula)
R> p <- 4 ##number of predictors
R> coords <- NYOzone.dat[,c("X.UTM","Y.UTM")]/1000
R> max.d <- max(iDist(coords))
R> starting <- list("beta"=rep(0,N.t*p), "phi"=rep(3/(0.5*max.d), N.t),
+                   "sigma.sq"=rep(2,N.t), "tau.sq"=rep(1, N.t),
+                   "sigma.eta"=diag(rep(0.01, p)))
R> tuning <- list("phi"=rep(2, N.t)) 
R> priors <- list("beta.0.Norm"=list(rep(0,p), diag(100000,p)),
+                 "phi.Unif"=list(rep(3/(0.9*max.d), N.t), rep(3/(0.05*max.d), N.t)),
+                 "sigma.sq.IG"=list(rep(2,N.t), rep(25,N.t)),
+                 "tau.sq.IG"=list(rep(2,N.t), rep(25,N.t)),
+                 "sigma.eta.IW"=list(2, diag(0.001,p)))
R> n.samples <- 5000
R> m.i <- spDynLM(mods, data=NYOzone.dat, coords=as.matrix(coords), 
+                 starting=starting, tuning=tuning, priors=priors, get.fitted=TRUE,
+                 cov.model="exponential", n.samples=n.samples, n.report=2500)
\end{Sinput}
\begin{Soutput}
----------------------------------------
	General model description
----------------------------------------
Model fit with 28 observations in 62 time steps.

Number of missing observations 117.

Number of covariates 4 (including intercept if specified).

Using the exponential spatial correlation model.

Number of MCMC samples 5000.

Priors and hyperpriors:
	beta normal:
	m_0:	0.000	0.000	0.000	0.000	
	Sigma_0:
	100000.000	0.000	0.000	0.000	
	0.000	100000.000	0.000	0.000	
	0.000	0.000	100000.000	0.000	
	0.000	0.000	0.000	100000.000	

	sigma.sq_t=1 IG hyperpriors shape=2.00000 and scale=25.00000
	tau.sq_t=1 IG hyperpriors shape=2.00000 and scale=25.00000
	phi_t=1 Unif hyperpriors a=0.00564 and b=0.10145
	---
	sigma.sq_t=2 IG hyperpriors shape=2.00000 and scale=25.00000
	tau.sq_t=2 IG hyperpriors shape=2.00000 and scale=25.00000
	phi_t=2 Unif hyperpriors a=0.00564 and b=0.10145
	---
	sigma.sq_t=3 IG hyperpriors shape=2.00000 and scale=25.00000
	tau.sq_t=3 IG hyperpriors shape=2.00000 and scale=25.00000
	phi_t=3 Unif hyperpriors a=0.00564 and b=0.10145
	---
        ...
	---
	sigma.sq_t=62 IG hyperpriors shape=2.00000 and scale=25.00000
	tau.sq_t=62 IG hyperpriors shape=2.00000 and scale=25.00000
	phi_t=62 Unif hyperpriors a=0.00564 and b=0.10145
	---
-------------------------------------------------
		Sampling
-------------------------------------------------
Sampled: 2499 of 5000, 49.98%
Report interval Mean Metrop. Acceptance rate: 49.05%
Overall Metrop. Acceptance rate: 49.07%
-------------------------------------------------
Sampled: 4999 of 5000, 99.98%
Report interval Mean Metrop. Acceptance rate: 49.48%
Overall Metrop. Acceptance rate: 49.28%
-------------------------------------------------
\end{Soutput}
\end{Schunk}

\begin{figure}[!ht]
\begin{center}
\includegraphics[width=15cm]{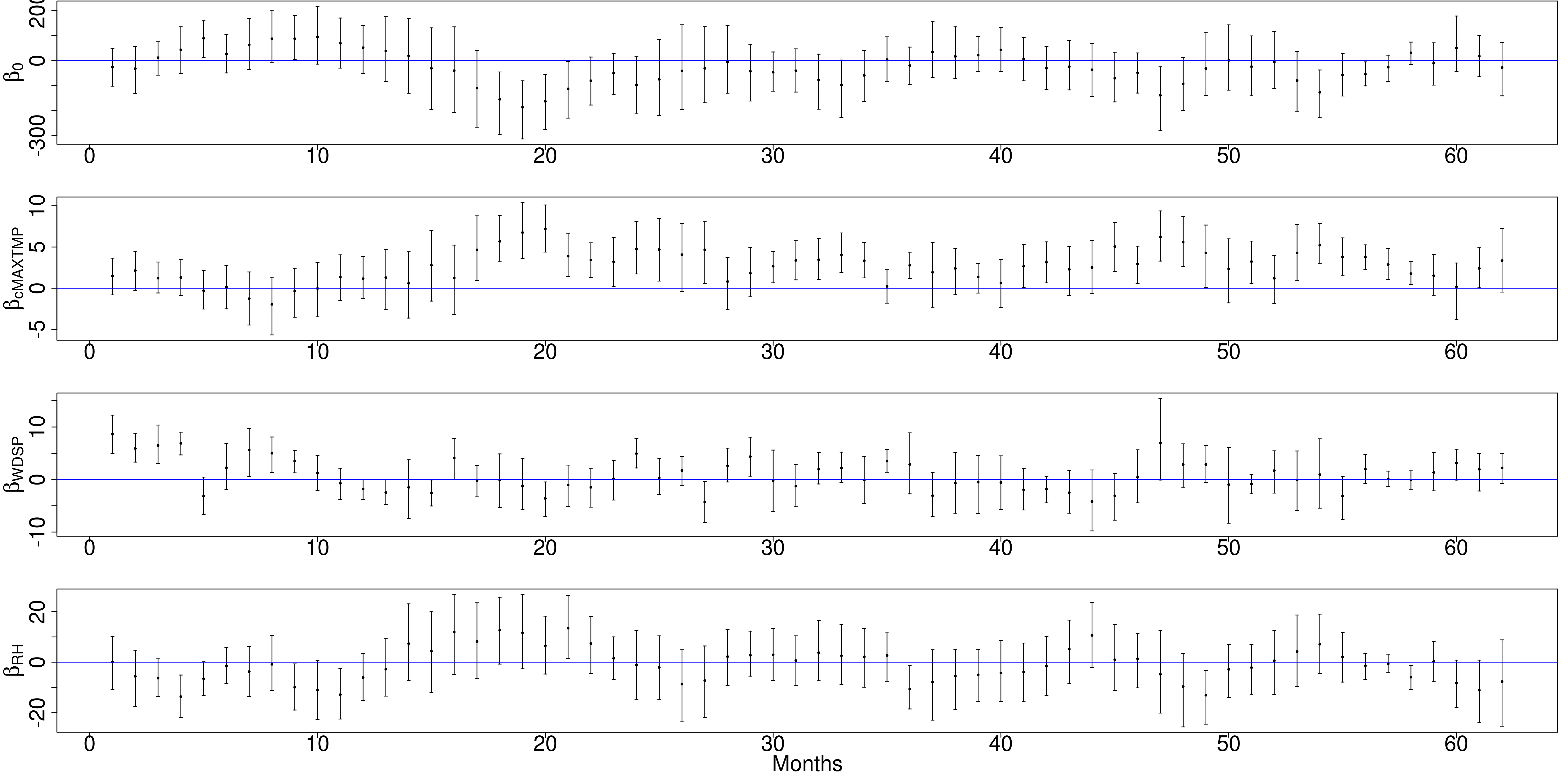}
\end{center}
\caption{Posterior distribution medians and 95\% credible intervals for model intercept and predictors.}
\label{fig:ozoneBeta}
\end{figure}

\begin{figure}[!ht]
\begin{center}
\includegraphics[width=15cm]{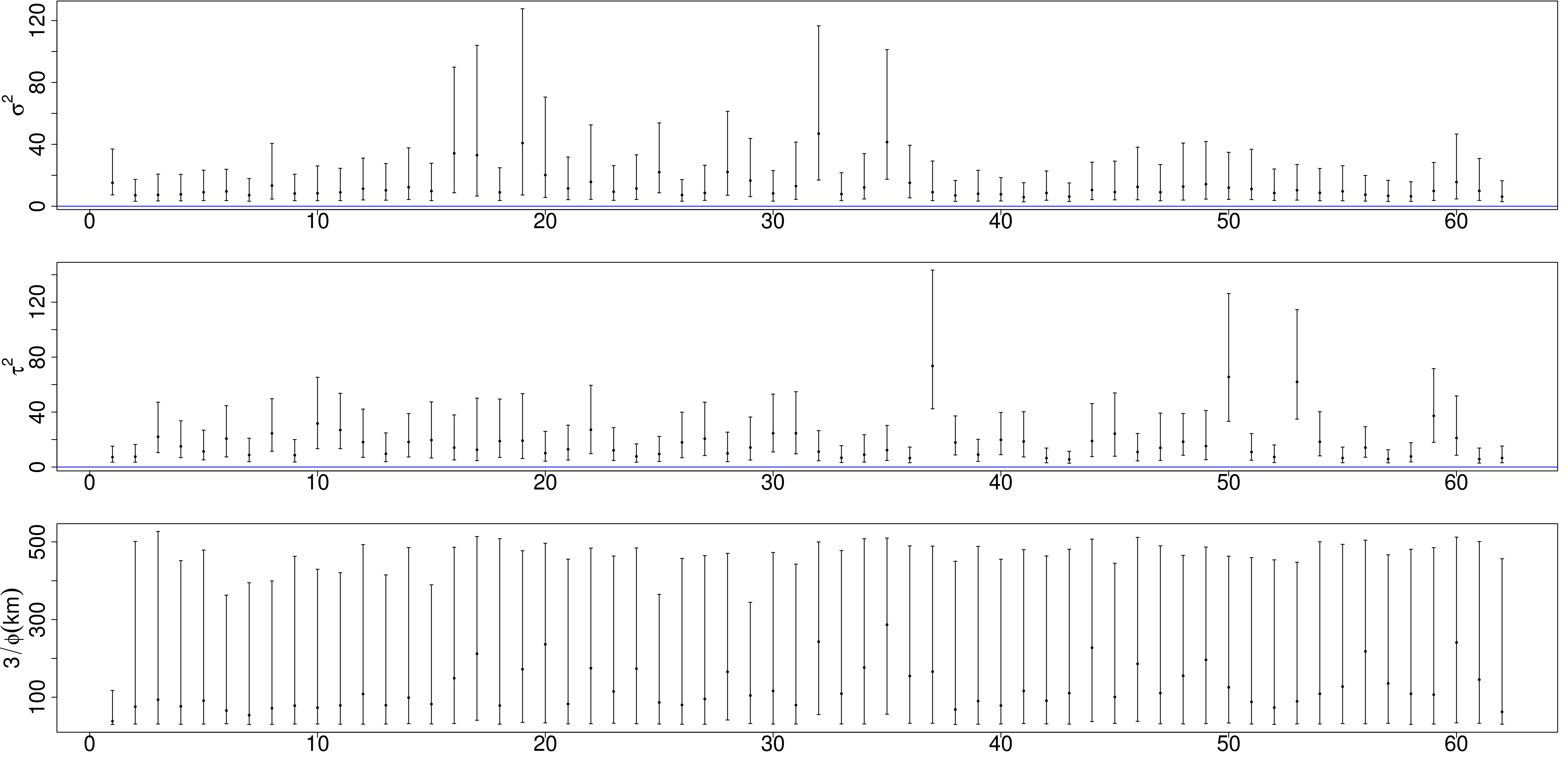}
\end{center}
\caption{Posterior distribution medians and 95\% credible intervals for $\btheta$ and $\tau^2$.}
\label{fig:ozoneTheta}
\end{figure}

Time series plots of parameters' posterior summary statistics are often useful for exploring the temporal evolution of the parameters. In the case of the regression coefficients, these plots describe the time-varying trend in the outcome and impact of covariates. For example, the sinusoidal pattern in the model intercept, $\beta_0$, seen in Figure~\ref{fig:ozoneBeta}, correlates strongly with both cMAXTMP, RM, and to a lesser degree with WDSP. With only a maximum of 28 observations within each time step, there is not much information to inform estimates of $\btheta$. As seen in Figure~\ref{fig:ozoneTheta}, this paucity of information is reflected in the imprecise CI's for the $\phi$'s and small deviations from the priors on $\sigma^2$ and $\tau^2$. There are, however, noticeable trends in the variance components over time.

\begin{figure}[!ht]
\begin{center}
\includegraphics[width=15cm]{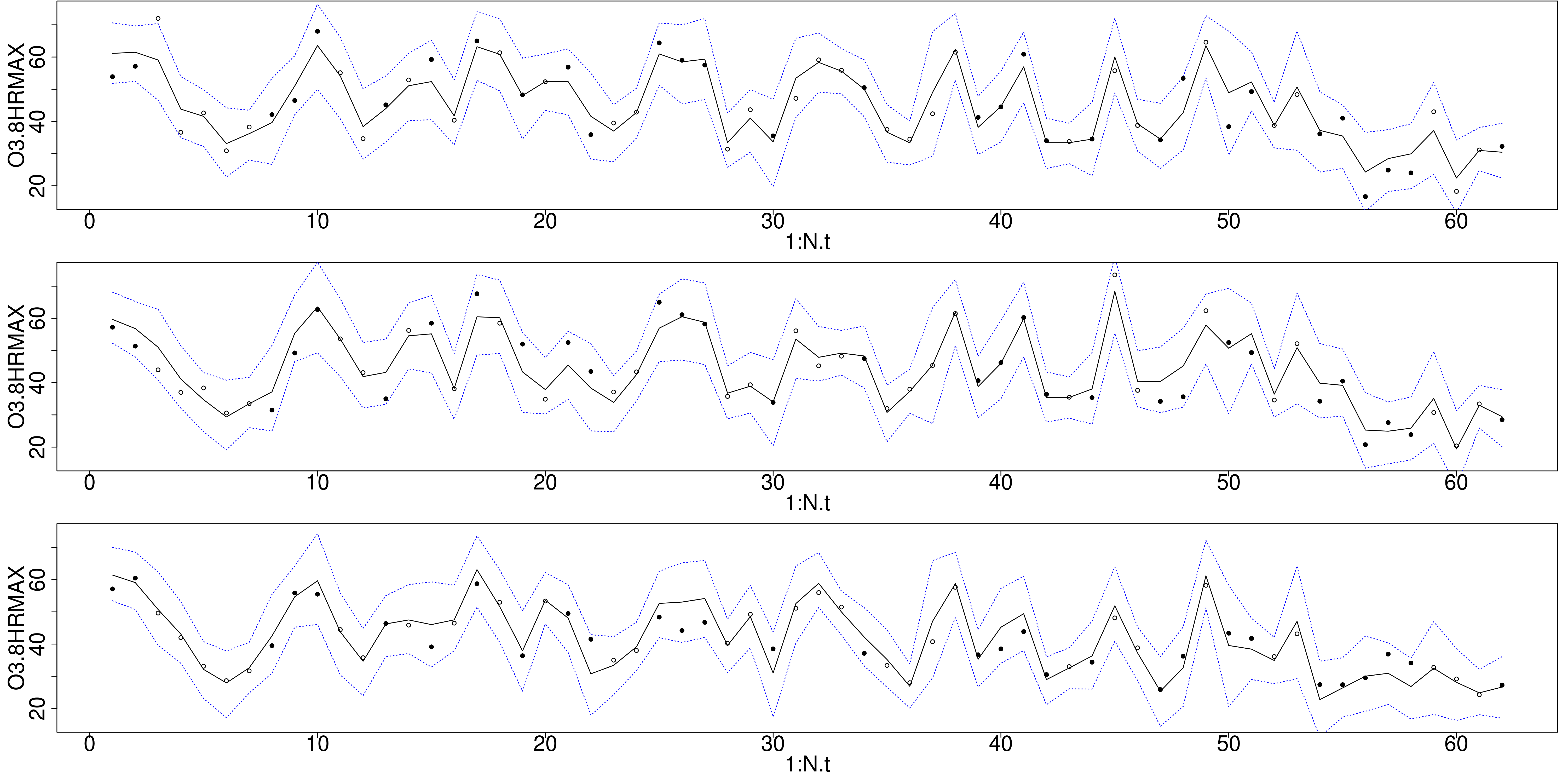}
\end{center}
\caption{Posterior predicted distribution medians and 95\% credible intervals, solid and dashed lines respectively, for three stations. Open circle symbols indicate those observations use for model parameter estimation and filled circle symbols indicate those observations withheld for validation.}
\label{fig:ozonePred}
\end{figure}

Figure~\ref{fig:ozonePred} shows the observed and predicted values for the three stations used for validation. Here, open circle symbols indicate those observations used for parameter estimation and filled circles identify holdout observations. The posterior predicted median and 95\% CIs are overlaid using solid and dashed lines, respectively. Three of the 36 holdout measurements fell outside of their 95\% predicted CI, a $\sim$92\% coverage rate. As noted in \cite{sahu2011}, there is a noticeable reduction in ozone levels in the last two weeks in August.

\section{Model choice}\label{Sec: Model_Choice}
The \texttt{spDiag} function provides several approaches to assessing model performance and subsequent comparison for \texttt{spLM}, \texttt{spMvLM}, \texttt{spGLM}, and \texttt{spMvGLM} objects. These include the popular Deviance Information Criterion \citep{spiegelhalter2002} as well as a measure of posterior predictive loss detailed in \cite{gelfand1998} and a scoring rule defined in \cite{gneiting2007}.

\section{Summary and future direction}
\pkg{spBayes} version 0.3-7 (CRAN 6/1/13), and subsequent versions, offers a complete reformulation and rewrite of core functions for efficient estimation of univariate and multivariate models for point-referenced data using MCMC. Substantial increase in computational efficiency and flexibility in model specification, compared earlier \pkg{spBayes} package versions, is the result of careful MCMC sampler formulation that focused on reducing parameter space and avoiding expensive matrix operations. In addition, all core functions provide \emph{predictive process} models able to accommodate large data sets that are being increasingly encountered in many fields. 

We are currently developing an efficient modeling framework and sampling algorithm to accommodate multivariate spatially misaligned data, i.e., settings where not all of the outcomes are observed at all locations, that will be added to the \code{spMvLM} and \code{spMvGLM} functions.  Prediction of these missing outcomes should borrow strength from the covariance among outcomes both within and across locations. In addition, we hope to add functions for non-stationary multivariate models such as those described in \cite{gelfand2004} and more recent predictive process versions we developed in \cite{guhaniyogi2013}. We will also continue developing \code{spDynLM} and helper functions. Ultimately, we would like to provide more flexible specifications of spatio-temporal dynamic models and allow them to accommodate non-Gaussian and multivariate outcomes.

\section*{Acknowledgments}
This work was supported by National Science Foundation grants DMS-1106609, EF-1137309, EF-1241868, and EF-1253225, as well as NASA Carbon Monitoring System grants.

\bibliography{mybib}{}

\end{document}